\documentclass[reqno,12pt]{amsart}
\usepackage{fullpage}

\usepackage{amssymb,amsmath,amsfonts,latexsym}
\usepackage{amsthm}

\usepackage{graphicx}
\usepackage{enumerate}
\usepackage{url}

\numberwithin{equation}{section}
\def\Rnum{\mathbb{R}}
\def\Mink#1{\Rnum^{#1,1}}

\def\pos{{\vec \gamma}}

\def\Im{{\rm Im}}
\def\Re{{\rm Re}}
\def\Hop{{\mathcal H}}
\def\Jop{{\mathcal J}}
\def\Rop{{\mathcal R}}
\def\Eop{{\mathcal E}}
\def\Iop{{\mathcal I}}
\def\Dop{{\mathcal D}}
\def\Cop{{\mathcal C}}

\def\X{{\mathrm X}}
\def\pr{{\rm pr}}
\def\Nop{{\mathcal N}}
\def\Qop{{\mathcal Q}}
\def\Sop{{\mathcal K}}
\def\Kop{{\mathcal S}}

\def\E{{\mathbf E}}
\def\U{{\mathbf U}}
\def\W{{\mathbf W}}
\def\H{{\mathbf H}}
\def\e{{\mathbf A}}

\def\mk{\mathfrak}

\def\t{{\rm T}}
\def\txtint{{\textstyle\int}}

\def\bmatr{\begin{pmatrix}}
\def\ematr{\end{pmatrix}}

\newtheorem{thm}{Theorem}{\bf}{\em}
\newtheorem{prop}{Proposition}{\bf}{\em}
{\bf}{\em}
{\bf}{\em}

\def\eqref#1{(\ref{#1})}

\def\thmref#1{Theorem~\ref{#1}}

\def\Ref#1{Ref.\cite{#1}}
\def\secref#1{Sec.~\ref{#1}}

\def\ie/{i.e.}
\def\eg/{e.g.}

\begin{document}

\title{Integrable systems from inelastic curve flows\\ in 2-- and 3-- dimensional Minkowski space}

\author{
Kivilcim Alkan$^{1,2}$ 
\lowercase{and}
Stephen C. Anco$^2$
\\
\\
\lowercase{\scshape{
${}^1$
Department of Mathematics\\ 
Izmir Institution of Technology\\
Izmir 35430 Turkey\\
kivilcimalkan@iyte.edu.tr}}\\
\lowercase{\scshape{
${}^2$
Department of Mathematics and Statistics\\
Brock University\\
St. Catharines, ON L2S3A1, Canada\\
sanco@brocku.ca}}
}

\begin{abstract}
Integrable systems are derived from inelastic flows of timelike, spacelike, 
and null curves in 2-- and 3-- dimensional Minkowski space. 
The derivation uses a Lorentzian version of a geometrical moving frame method 
which is known to yield the modified Korteveg-de Vries (mKdV) equation 
and the nonlinear Schr\"odinger (NLS) equation 
in 2-- and 3-- dimensional Euclidean space, respectively. 
In 2--dimensional Minkowski space, 
timelike/spacelike inelastic curve flows are shown to yield 
the defocusing mKdV equation and its bi-Hamiltonian integrability structure,
while inelastic null curve flows are shown to give rise to Burgers' equation
and its symmetry integrability structure. 
In 3--dimensional Minkowski space, 
the complex defocusing mKdV equation and the NLS equation 
along with their bi-Hamiltonian integrability structures 
are obtained from timelike inelastic curve flows,
whereas spacelike inelastic curve flows yield an interesting variant of 
these two integrable equations in which complex numbers are replaced by 
hyperbolic (split-complex) numbers.
\end{abstract}

\maketitle

\section{\large Introduction}

There is an interesting geometric relationship between 
integrable systems and non-stretching (inelastic) geometrical flows of curves 
in various geometric spaces. 
For example, in the Euclidean plane, 
inelastic curve flows yield \cite{GolPet} the focusing mKdV equation 
$u_t = u_{xxx} + \tfrac{3}{2}u^2u_x$ 
where $x$ is the arclength of the curve
and $u$ is the curvature invariant of the curve,  
with the geometrical motion of the curve 
being given by the vortex patch equation. 
In Euclidean space, 
inelastic curve flows whose geometrical motion is given by 
the vortex filament equation \cite{Has} 
and its axial generalization \cite{FukMiy}
respectively yield \cite{Has,Lam,MarSanWan,FukMiy,AncMyr}
the focusing NLS equation 
$-iu_t = u_{xx} + \tfrac{1}{2}|u|^2u$ 
and the focusing complex mKdV equation 
$u_t = u_{xxx} + \tfrac{3}{2}|u|^2u_x$. 
Here $u=\kappa \exp(i\int \tau dx)$ is the Hasimoto variable \cite{Has}
which is a covariant of the curve 
defined in terms of the curvature and torsion invariants $\kappa$ and $\tau$ 
up to arbitrary (constant) phase rotations $u\rightarrow e^{i\phi} u$. 

The derivation of these integrable systems from the underlying curve flows
is based on a moving frame method that uses 
a Frenet frame \cite{Gug} in the case of curves in the Euclidean plane
and a parallel frame \cite{Bis} in the case of curves in Euclidean space. 
Parallel frames differ from a Frenet frame by a gauge transformation 
given by a rotation of the two normal vectors 
(in the normal plane of the curve) 
such that, at each point, 
their derivative along the curve is purely tangential. 
In both the case of the Euclidean plane and Euclidean space, 
the components of the Cartan matrix of the frame define the flow variable $u$,
and the Cartan structure equations of the frame yield 
a Lax pair as well as a pair of Hamiltonian operators,
which provide the integrability structure of the flows on $u$. 
In addition, in the case of space curves, 
the $U(1)$ phase-rotation invariance of the flows on $u$
geometrically corresponds to the fact that a parallel frame 
is determined by a curve only up to the action of 
arbitrary rigid $SO(2)$ rotations on the normal vectors in the frame, 
whereas a Frenet frame is determined uniquely by a curve.

In this paper, 
integrable systems are derived in an analogous way from 
inelastic curve flows in the Minkowski plane and in Minkowski space
\cite{ONei}. 
Several new results are obtained by considering 
timelike curves, spacelike curves, and null curves. 
The paper is organized as follows. 

In \secref{2dim}, 
a Lorentzian version of a Frenet frame is applied to 
inelastic curve flows in the Minkowski plane. 
Geometrical inelastic flows of timelike and spacelike curves 
are found to yield the defocusing mKdV equation 
and its bi-Hamiltonian structure. 
In contrast, for null curves, a Frenet frame does not exist 
and instead a null frame is introduced. 
Geometrical inelastic flows of null curves are shown to yield 
Burger's equation and the Cole-Hopf transformation 
under which Burger's equation is mapped into the heat equation. 
This geometric realization of Burger's equation is new. 
(See \Ref{FujKur} for related work.)
Frame formulations of inelastic flows of timelike, spacelike, and null curves 
has appeared previously in \Ref{Gur}.

In \secref{3dimtimelike}, 
starting from a Frenet frame, 
a Lorentzian counterpart of a parallel frame is introduced
for timelike curves in 3-dimensional Minkowski space. 
The normal plane of a timelike curve is spacelike, 
and so a Lorentzian parallel frame is determined by a curve 
up to the action of rigid $SO(2)$ rotations,
similarly to the case of Euclidean space. 
Consequently, the Cartan matrix of this frame yields 
the same complex-valued Hasimoto variable as in the case of Euclidean space. 
For timelike curves undergoing an inelastic geometrical flow
given by a timelike version of the vortex filament equation
and its axial generalization, 
the Cartan structure equations of the Lorentzian parallel frame 
are shown to yield the defocusing NLS equation 
and the defocusing complex mKdV equation 
along with their bi-Hamiltonian integrability structure. 
A similar derivation of the defocusing NLS equation 
has appeared in \Ref{DinIno} without, however, 
deriving the bi-Hamiltonian structure of this equation
or considering the defocusing complex mKdV equation. 
Parallel frames for non-null curves have been considered previously in \Ref{OzdErg}. 

In \secref{3dimspacelike}, 
spacelike curves in Minkowski space are considered. 
Because the normal plane of a spacelike curve is timelike, 
two different cases arise depending on whether 
the principal normal vector of the curve is non-null or null. 

When the principal normal vector of a spacelike curve is non-null, 
a Frenet frame exists which is used to define a Lorentzian parallel frame
by a gauge transformation given by a boost (hyperbolic rotation) 
of the two normal vectors. 
Consequently, the resulting frame is determined by the curve only 
up to the action of rigid $SO(1,1)$ boosts.
These boosts comprise a group of hyperbolic rotations,
and the Hasimoto variable arising from the Cartan matrix of this frame 
is no longer a complex variable but instead is a hyperbolic generalization 
\cite{splitcomplex}
based on the split-complex numbers defined by $j^2 = 1$ and $\bar{j} =-j$. 
For spacelike curves undergoing an inelastic geometrical flow
given by a spacelike generalization of the vortex filament equation 
and its axial generalization, 
the Cartan structure equations of a non-null parallel frame 
are shown to yield variants of the defocusing NLS equation 
and the defocusing complex mKdV equation, 
along with their bi-Hamiltonian integrability structure, 
in which $i$ is replaced by $j$. 
Such integrable systems have been derived previously in the literature
\cite{BraHay} by purely algebraic methods. 
Our results provide an interesting geometric realization for these systems. 
A special case of the $SO(1,1)$-invariant complex mKdV equation 
has been derived in \Ref{DinWanWan} using similar geometrical methods, 
and a system equivalent to the $SO(1,1)$-invariant NLS equation appears in \Ref{DinIno} in a less geometric form, without any results on the bi-Hamiltonian structure of these equations. 

When a spacelike curve has a null principal normal vector, 
a complete Frenet frame does not exist. 
Instead a null frame is introduced for the normal plane. 
Then a Lorentzian parallel frame is defined by a gauge transformation 
such that the derivative of the pair of null normal vectors in the frame
is purely tangential to the curve. 
The gauge transformation acts as a scaling on these null normal vectors,
and consequently the resulting frame is determined by the curve only 
up to the action of rigid scalings
(where the two null vectors scale reciprocally to each other). 
Because the principal normal vector of the curve is constrained to be null,
the Cartan matrix has one fewer component than in the case when 
the principal normal vector is non-null. 
Interestingly, the Cartan structure equations are found to yield 
Burgers' equation and the KdV equation,
as well as the heat equation and the Airy equation. 
These results are new. 
The corresponding geometrical inelastic curve flows are shown to be given by 
variants of a heat map equation and an Airy map equation. 

Related work on null curve flows in 3-- and 4-- dimensional Minkowski space 
appears in \Ref{Gur,MusNic,Nak}. 

Some concluding remarks are made in \secref{conclude}.

\section{Curve flows and integrable systems in $\Mink{1}$}
\label{2dim}

The Minkowski plane is a 2-dimensional vector space $\Mink{1}$
equipped with a Lorentz-signature metric $\eta$
and a compatible volume form $\epsilon$.  
A vector $\vec v\in \Mink{1}$ is respectively 
{\em timelike}, {\em spacelike}, or {\em null} if 
its Minkowski norm $\eta(\vec v,\vec v)$ is negative, positive, or zero. 
The set of all null vectors spans a pair of 1-dimensional null lines
through the origin in $\Mink{1}$, called the {\em lightcone}. 

Up to a sign, the volume form is determined from the metric by the relation 
\begin{equation}
\epsilon(\vec v,\vec w)^2
=\eta(\vec v,\vec w)^2 -\eta(\vec v,\vec v)\eta(\vec w,\vec w)
\end{equation}
holding for any vectors $\vec v,\vec w \in \Mink{1}$.
Associated to the metric and the volume form is 
the Hodge dual operator $*$ defined by 
\begin{equation}\label{dual}
\eta(*\vec v,\vec w) = \epsilon(\vec v,\vec w)
\end{equation}
which has the properties
\begin{equation}\label{dual1}
*(*\vec v) = \vec v,
\quad
\eta(*\vec v,\vec v)=0
\end{equation}
and 
\begin{equation}\label{dual2}
\eta(*\vec v,\vec w)=-\eta(*\vec w,\vec v),
\quad
\eta(*\vec v,*\vec w)=-\eta(\vec v,\vec w)
\end{equation}
holding for any vectors $\vec v,\vec w \in \Mink{1}$.

In inertial coordinates $ X^{i}=(X^{0},X^{1})$, 
the Minkowski metric has the components
$\eta_{ij}=\bmatr -1 & 0\\ 0 &  1\ematr$ 
and the volume tensor has the components
$\epsilon_{ij}=\bmatr 0 & 1\\ -1 &  0\ematr$,
where $X^0$ is a timelike coordinate and $X^1$ is a spacelike coordinate. 
The Hodge dual operator $*$ is given by the associated tensor 
$\epsilon_i{}^j=\epsilon_{ik}\eta^{jk}$.  

The isometries of the Minkowski plane consist of time and space translations
and boosts. 
These transformations comprise the 3-dimensional Poincare group 
$ISO(1,1)\simeq SO(1,1)\ltimes \Rnum^2$
(also called the inhomogeneous Lorentz group).

\subsection{Timelike/spacelike curves and Frenet frames}
\label{2dimnonnullcase}
Let $\pos(x)=(X^0(x),X^1(x))$ be a curve in $\Mink{1}$ such that
\begin{equation}
\eta(\pos_{x},\pos_{x})\neq 0
\label{2dimsignT}
\end{equation}
at every point $x$ on the curve. 
Such a curve is timelike if $\eta(\pos_{x},\pos_{x})<0$ 
or spacelike if  $\eta(\pos_{x},\pos_{x})>0$.
In both cases we hereafter choose $x$ to be the proper-time or proper-distance
arclength parameter defined by 
\begin{equation}
\sqrt{|\eta(\pos_{x},\pos_{x})|}=1 . 
\label{2dimarclength}
\end{equation}

From relation \eqref{dual1}, 
we note that the vectors $\pos_x$ and $*\pos_x$ are orthogonal 
in the Minkowski metric. 
Since $\pos_x$ is assumed to be non-null, 
then this implies $*\pos_x$ is also non-null. 
Hence, for a timelike/spacelike curve $\pos(x)$ in $\Mink{1}$,
this pair of vectors can be used to define 
a Lorentzian version of a Frenet frame \cite{Gur}
\begin{subequations}\label{2dimfrenetframe}
\begin{align}
& e_\parallel =\pos_x=T, 
\quad
\text{unit timelike/spacelike tangent vector} 
\label{2dimT}
\\   
& e_\perp=*\pos_x=*T=N,
\quad 
\text{unit spacelike/timelike normal vector} 
\label{2dimN}
\end{align}
\end{subequations}
satisfying 
\begin{gather}
\eta(e_\parallel,e_\parallel)=\mp 1, 
\quad
\eta(e_\perp,e_\perp)=\pm 1,
\label{frenet1}
\\
\eta(e_\parallel,e_\perp)=0 . 
\label{frenet2}
\end{gather}

The Frenet equations for this frame \eqref{2dimfrenetframe} are 
easily derived as follows. 
First, from the $x$-derivative of equation \eqref{frenet1}, 
we have $\eta(e_\parallel{}_x,e_\parallel)=0$ 
and $\eta(e_\perp{}_x,e_\perp)=0$,
which implies $e_\parallel{}_x = u_1 e_\perp$ and $e_\perp{}_x = u_2 e_\parallel$
for some functions $u_1(x)$ and $u_2(x)$. 
Next, from the $x$-derivative of equation \eqref{frenet2}, 
we have $\eta(e_\parallel{}_x,e_\perp)+\eta(e_\perp{}_x,e_\parallel)=0$. 
Now substituting the previous expressions for the $x$-derivatives of 
$e_\parallel$ and $e_\perp$, 
and then using the norm relations \eqref{frenet1}, 
we find $0=u_1\eta(e_\perp,e_\perp) + u_2\eta(e_\parallel,e_\parallel)= \pm(u_1-u_2)$. 
This yields $u_1=u_2=u$ in both cases. 
Thus the Frenet equations are given by 
\begin{equation}
\bmatr e_{\parallel}\\e_{\perp} \ematr_{x}
=\bmatr 0&u\\u&0 \ematr \bmatr e_{\parallel}\\e_{\perp} \ematr
\label{2dimfreneteqs}
\end{equation}
where $u$ geometrically represents 
the Lorentzian curvature invariant of the curve $\pos(x)$,
and where the Cartan matrix $\bmatr 0&u\\u&0 \ematr$
belongs to the Lie algebra $\mathfrak{so}(1,1)$ of the 
$SO(1,1)$ group of boost isometries in $\Mink{1}$.
 
We now consider curve flows $\pos(t,x)$ that locally preserve 
both the timelike/spacelike signature \eqref{2dimsignT} of the curve 
and the proper time/distance normalization \eqref{2dimarclength} of 
the arclength parameter. 
Such flows are called {\em inelastic}
and are specified by a flow vector 
\begin{equation}\label{2dimflow}
\pos_t=h_\parallel e_\parallel + h_\perp e_\perp
\end{equation}
expressed in terms of tangential and normal components 
in the Frenet frame \eqref{2dimfrenetframe}. 
The Frenet frame itself will be carried by the flow, 
according to the equations 
\begin{equation}
\bmatr e_{\parallel}\\e_{\perp} \ematr_{t}
=\bmatr 0&\omega \\\omega &0 \ematr\bmatr e_{\parallel}\\e_{\perp} \ematr
\label{2dimfloweqs}
\end{equation}
which arise from equations \eqref{frenet1}--\eqref{frenet2} 
similarly to the derivation of the Frenet equations \eqref{2dimfreneteqs}. 
Here $\omega$ geometrically represents a Lorentzian invariant of the curve flow,
and the Cartan matrix $\bmatr 0&\omega\\ \omega &0 \ematr$
belongs to the Lie algebra $\mathfrak{so}(1,1)$. 

The Frenet equations \eqref{2dimfreneteqs} and the flow equations \eqref{2dimfloweqs} 
are related by the compatibility condition that the mixed $t,x$-derivatives 
of the Frenet frame \eqref{2dimfrenetframe} are equal.
Since the Cartan matrices commute,
this yields 
\begin{equation}
u_t=\omega_x . 
\label{2dimstructeq1}
\end{equation}
Likewise, 
the tangent vector \eqref{2dimT} and the flow vector \eqref{2dimflow}
of the curve are related by the compatibility condition that 
the mixed $t,x$-derivatives of $\pos(t,x)$ are equal, 
which gives
\begin{equation}
h_\perp{}_xe_\perp+h_\perp e_\perp{}_x+h_\parallel{}_x e_\parallel+h_\parallel e_\parallel{}_x
=e_\parallel{}_t .  
\end{equation}
After substituting the Frenet equations \eqref{2dimfreneteqs} 
and the flow equations \eqref{2dimfloweqs} 
for the derivatives of $e_{\parallel}$ and $e_{\perp}$, 
we obtain 
\begin{equation}
e_\parallel{(h_\parallel{}_x+uh_\perp)}+e_\perp{(h_\perp{}_x+uh_\parallel-\omega)}=0.
\end{equation}
Hence, from the vanishing of the coefficients of the frame vectors 
$e_{\parallel}$ and $e_{\perp}$, 
we find 
\begin{gather}
\omega=h_\perp{}_x+uh_\parallel , 
\label{2dimstructeq2perp}
\\
h_\parallel{}_x=-uh_\perp . 
\label{2dimstructeq2par}
\end{gather}

These compatibility equations 
\eqref{2dimstructeq2perp}--\eqref{2dimstructeq2par} and \eqref{2dimstructeq1}
are the Cartan structure equations of the Frenet frame. 
They have a natural operator formulation as follows. 
We use equation \eqref{2dimstructeq2par} to express 
$h_\parallel=-D_{x}^{-1}(u{h_\perp})$, 
so then \eqref{2dimstructeq2perp} becomes 
\begin{equation}\label{2dimstructeq2} 
\omega=h_\perp{}_x-uD_{x}^{-1}(u{h_\perp}) . 
\end{equation}
Hence, the Cartan structure equations reduce to 
the system \eqref{2dimstructeq2} and \eqref{2dimstructeq1}. 

Recall that an operator $\Hop$ is Hamiltonian iff 
it defines an associated Poisson bracket
\begin{equation}\label{realpoisson}
\{\mk{H},\mk{G} \}_{\Hop} =
\int (\delta \mk{H}/\delta u) \Dop( \delta \mk{G}/\delta u) dx
\end{equation}
obeying skew-symmetry 
$\{\mk{H},\mk{G} \}_{\Hop} + \{\mk{G},\mk{H} \}_{\Hop} =0$
and the Jacobi identity 
$\{\{\mk{H},\mk{G} \}_{\Hop},\mk{F}\}_{\Hop} + \text{cyclic } =0$
for all functionals $\mk{H}$, $\mk{G}$, $\mk{F}$ 
on the $x$-jet space $J^\infty$ of the variable $u$,
where $\delta /\delta u$ denotes the standard Euler operator. 
The formal inverse of a Hamiltonian operator defines a symplectic operator. 
Compatibility of a Hamiltonian operator $\Hop$ and a symplectic operator $\Jop$
is the statement that every linear combination 
$c_{1}\Hop+c_{2}\Jop^{-1}$ is a Hamiltonian operator, 
or equivalently that $c_{1}\Hop^{-1}+c_{2}\Jop$ is a symplectic operator. 

\begin{thm}
\label{2dimHJthm}
For timelike or spacelike inelastic curve flows in $\Mink{1}$, 
the curve invariant $u$ satisfies the system 
\begin{gather}
u_t=\Hop(\omega), 
\quad 
\Hop=D_x
\label{2dimH}
\\
\omega=\Jop(h_\perp), 
\quad 
\Jop =D_x -uD_{x}^{-1}u
\label{2dimJ}
\end{gather}
where $\Hop$ and $\Jop$ are a pair of 
compatible Hamiltonian and symplectic operators.
Composition of these operators yields the evolution equation 
\begin{equation}\label{uflow}
u_t=\Rop(h_\perp)
\end{equation}
for $u$ in terms of the normal component $h_\perp$ of the flow, 
where 
\begin{equation}
\Rop=\Hop\Jop= D_x{}^2 -u^2 -u_xD_x^{-1}u 
\end{equation}
is a hereditary recursion operator. 
\end{thm}

This theorem is a counterpart of a similar result 
(see e.g. \cite{SanWan,Anc08})
known for inelastic curve flows in the Euclidean plane 
and can be proved by the same methods. 
In particular, compared to the Euclidean case, 
the Hamiltonian operator \eqref{2dimH} is exactly the same 
while the symplectic operator \eqref{2dimJ} differs only by 
the sign of the nonlocal term. 

As a consequence of the compatibility of this pair of operators 
\eqref{2dimH} and \eqref{2dimJ}, 
their obvious invariance under $x$-translations can be used to derive
a hierarchy of flows starting from a root flow 
$u_t=u_x$ corresponding to the generator of $x$-translations on $u$, 
due to a general theorem of Magri \cite{Mag}. 
This leads to the following main result. 

\begin{thm}
\label{2dimhierarchy}
There is a hierarchy of integrable bi-Hamiltonian flows on $u(t,x)$
given by
\begin{equation}
\begin{aligned}
u_t=\Rop^n(u_x) & 
= \Hop(\delta \mk{H}^{(n)}/\delta u) , 
\quad 
n=0,1,2,\ldots 
\\& 
=\Eop(\delta \mathfrak{H}^{(n-1)}/\delta u), 
\quad 
n=1,2,\ldots 
\end{aligned}
\label{2dimbihamil}
\end{equation}
(called the {\em $+n$ flow})
in terms of Hamiltonians $\mk{H}^{(n)}=\int H^{(n)} dx$
where 
\begin{equation}
\Hop= D_x,
\quad
\Eop = \Rop\Hop= D_x{}^3 -u^2D_x -u_xD_x^{-1}uD_x 
\end{equation}
are compatible Hamiltonian operators, 
and where 
\begin{equation}
H^{(n)}=
(1+2n)^{-1} D^{-1}_x( u\Rop^n(u_x) ), 
\quad 
n=0,1,2,\ldots
\label{2dimHam}
\end{equation}
are local Hamiltonian densities. 
\end{thm}

The expression for the Hamiltonians \eqref{2dimHam} in this theorem 
arises from a scaling formula derived in \Ref{Anc03}. 

The $+1$ flow in the hierarchy \eqref{2dimbihamil} is explicitly given by 
\begin{equation}\label{2dimmkdv}
u_t=u_{xxx}-\tfrac{3}{2}u^2 u_x 
\end{equation}
which is the defocusing mKdV equation. 
It has the explicit bi-Hamiltonian structure
\begin{equation}\label{2dimmkdvhamil}
u_t = \Hop(\delta \mk{H}^{(1)}/\delta u) = \Eop(\delta\mk{H}^{(0)}/\delta u) 
\end{equation}
in terms of the Hamiltonian densities 
\begin{equation}\label{2dimmkdvH}
H^{(0)} = \tfrac{1}{2} u^2,
\quad
H^{(1)} = -\tfrac{1}{2} (u_x^2 + \tfrac{1}{4} u^4)
\end{equation}
(modulo trivial densities given by total $x$-derivatives). 

The entire hierarchy \eqref{2dimbihamil} of flows corresponds to 
a hierarchy of commuting vector fields 
\begin{equation}\label{2dimsymms}
\X^{(n)} = \Rop^n(u_x)\partial_u,
\quad
n=0,1,2,\ldots
\end{equation}
with the root vector field being the generator of $x$-translations, 
$\X^{(0)} = u_x\partial_u$. 
Recall that a vector field is Hamiltonian iff 
its prolongation on the $x$-jet space $J^\infty$ satisfies
$\pr\X\mk{G} = \{\mk{G},\mk{H} \}_{\Hop}$ for some functional $\mk{H}$,
where $\mk{G}$ is an arbitrary functional. 
\thmref{2dimhierarchy} shows that all of the vector fields 
in this hierarchy are Hamiltonian, as given by 
$\pr\X^{(n)}\mk{G} = \{\mk{G},\mk{H}^{(n)}\}_{\Hop}$,
where $\mk{H}^{(n)}=\int H^{(n)}dx$ is the functional 
with the Hamiltonian density \eqref{2dimHam}.
Additionally, each vector field except for $\X^{(0)} = u_x\partial_u$ 
is bi-Hamiltonian, 
due to $\pr\X^{(n)}\mk{G} = \{\mk{G},\mk{H}^{(n-1)}\}_{\Eop}$ for all $n\geq 1$. 
Since the hierarchy for all $n\geq 0$ is commuting, 
every Hamiltonian vector field \eqref{2dimsymms} 
is the generator of a symmetry 
for the defocusing mKdV equation \eqref{2dimmkdv}, 
and every associated Hamiltonian \eqref{2dimHam} 
is a conserved density 
for the defocusing mKdV equation \eqref{2dimmkdv}. 

\begin{prop}
Each flow in the hierarchy \eqref{2dimbihamil} 
determines an inelastic timelike/spacelike curve flow 
\begin{equation}\label{2dimcurveflow}
\pos_t=h_\parallel^{(n)}T + h_\perp^{(n)} N,
\quad
n=1,2,\ldots
\end{equation}
whose tangential and normal components are given by 
\begin{equation}\label{2dimcurveflowh}
h_\parallel^{(n)}= (2n-1)H^{(n-1)},
\quad
h_\perp^{(n)}= \Rop^{n-1}(u_x),
\quad
n=1,2,\ldots
\end{equation}
where 
\begin{equation}
T=\pos_x,
\quad
N=*\pos_x . 
\end{equation}
The components \eqref{2dimcurveflowh} of these flows 
are functions of the curve invariant $u$ and its $x$-derivatives, 
so thus the equation of motion \eqref{2dimcurveflow} are invariant under 
the isometry group $ISO(1,1)$ of the Minkowski plane. 
\end{prop}

Each equation of motion \eqref{2dimcurveflow} 
describes a geometric non-stretching motion of the curve.
The curve flow $n=1$ 
corresponding to the defocusing mKdV equation \eqref{2dimmkdv}
is given by 
\begin{equation}\label{2dimmkdvcurveflow}
\pos_{t}=-\tfrac{1}{2}u^2 T +u_x N . 
\end{equation}
This equation of motion is a Lorentzian version of the vortex patch equation. 
It can be expressed entirely in terms of 
$\pos_x$, $*\pos_x$, and their $x$-derivatives
through the Frenet equations \eqref{2dimfreneteqs} as follows.
We have 
\begin{equation}
u=\pm\eta(e_\parallel{}_x,e_\perp) = \pm \eta(*\pos_x,\pos_{xx}),
\quad
u^2=\pm\eta(e_\parallel{}_x,e_\parallel{}_x) = \pm \eta(\pos_{xx},\pos_{xx}) 
\end{equation}
and hence
\begin{equation}
u_x=\pm \eta(*\pos_x,\pos_{xxx})
\end{equation}
using the relation \eqref{dual1}. 
Thus the equation of motion \eqref{2dimmkdvcurveflow} becomes 
\begin{equation}
\pos_{t}=-\tfrac{1}{2}\eta(*\pos_x,\pos_{xx})^2 \pos_x \pm \eta(*\pos_x,\pos_{xxx}) {*\pos_x} ,
\quad 
\eta(\pos_x,\pos_x)=\mp 1 . 
\end{equation} 
We can simplify this equation of motion further by using the frame expansion 
\begin{equation}
\vec v = \pm( \eta(\vec v,e_\perp)e_\perp - \eta(\vec v,e_\parallel)e_\parallel )
\end{equation} 
holding for any vector $\vec v\in \Mink{1}$. 
This yields 
\begin{equation}
\pos_{t}=\mp\tfrac{3}{2}\eta(\pos_{xx},\pos_{xx}) \pos_x + \pos_{xxx}, 
\quad 
\eta(\pos_x,\pos_x)=\mp 1
\end{equation} 
which is a Lorentzian version of the non-stretching mKdV map equation
\cite{Anc06}, in the Minkowski plane.

\subsection{Null curves and null frames}
\label{2dimnullcase}
Let $\pos(x)=(X^0(x),X^1(x))$ be a null curve in $\Mink{1}$, 
satisfying
\begin{equation}\label{2dimnull}
\eta(\pos_{x},\pos_{x})= 0
\end{equation}
at every point $x$ on the curve.
Here, in contrast to the case of non-null curves, 
the parameter $x$ is arbitrary 
and cannot be normalized solely in terms of $\pos_x$. 

Moreover, 
the vectors $\pos_x$ and $*\pos_x$ are now parallel,
since we have 
$0=\eta(\pos_x,\pos_x)=-\eta(*\pos_x,*\pos_x)=\epsilon(*\pos_x,\pos_x)$, 
from the relations \eqref{dual} and \eqref{dual2},
where antisymmetry of $\epsilon$ directly implies $*\pos_x=a \pos_x$
for some function $a(x)$. 
Similarly, the vectors $\pos_x$ and $\pos_{xx}$ are parallel,
since the $x$-derivative of $\eta(\pos_x,\pos_x)=0$ yields 
$\eta(\pos_x,\pos_{xx})=0$ from which we obtain
$\epsilon(*\pos_x,\pos_{xx}) = -\eta(\pos_x,\pos_{xx})=0$ 
through the relations \eqref{dual} and \eqref{dual2}. 
Antisymmetry of $\epsilon$ then implies $\pos_{xx}=b {*\pos_x}$
for some function $b(x)$, 
and thus $\pos_{xx}$ is parallel to $*\pos_x$ 
which itself is parallel to $\pos_x$. 

Hence, 
for defining a frame, 
we need another vector, linearly independent of $\pos_x$. 
One natural choice is to use a null vector 
on the opposite side of the lightcone. 
If $\vec v$ is a null vector in $\Mink{1}$, 
let $\Nop$ be a linear map that produces a null vector $\Nop(\vec v)$
such that $\eta(\Nop(\vec v),\vec v) = -1$. 
Note this implies $\vec v+\Nop(\vec v)$ is a timelike vector, since 
$\eta(\vec v+\Nop(\vec v),\vec v+\Nop(\vec v)) 
= 2\eta(\vec v,\Nop(\vec v))= -2$.
Hence the null vectors $\vec v$ and $\Nop(\vec v)$ 
are spatial reflections of each other with respect to this timelike vector 
in $\Mink{1}$, with the reflection being given by 
$\vec v - \tfrac{1}{2}(\vec v+\Nop(\vec v)) 
= \tfrac{1}{2}(\vec v-\Nop(\vec v)) 
= -(\Nop(\vec v) - \tfrac{1}{2}(\vec v+\Nop(\vec v)))$
where $\vec v-\Nop(\vec v)$ is a spacelike vector
orthogonal to $\vec v+\Nop(\vec v)$.
Also note that a change in the normalization $\eta(\Nop(\vec v),\vec v) = -1$ 
only produces a scaling of the null vector $\Nop(\vec v)$. 

A null frame for a null curve $\pos(x)$ in $\Mink{1}$ 
can then be defined by 
\begin{subequations}\label{2dimnullframe}
\begin{align}
& e_{+}=\pos_x=T, 
\quad
\text{null tangent vector}
\label{2dimnullT}
\\
& e_{-} = \Nop(\pos_x) = \Nop(T), 
\quad
\text{null opposite vector}
\label{2dimnullopp}
\end{align}
\end{subequations}
with 
\begin{gather}
\eta(e_{+},e_{+})=\eta(e_{-},e_{-})=0
\label{2dimnull1}
\\
\eta(e_{+},e_{-})=-1
\label{2dimnull2}
\end{gather}
where these properties \eqref{2dimnull1}--\eqref{2dimnull2} uniquely determine 
$e_{-}$ when $e_{+}$ is given. 

The Frenet equations for this frame \eqref{2dimnullframe} are 
easily derived by the same steps used in the case of a Frenet frame. 
First, the $x$-derivative of equation \eqref{2dimnull1} yields
$\eta(e_{+}{}_x,e_{+})=0$ 
and $\eta(e_{-}{}_x,e_{-})=0$. 
This implies $e_{\pm}{}_x = u_{\pm} e_{\pm}$ 
for some functions $u_{\pm}(x)$. 
Next, the $x$-derivative of equation \eqref{2dimnull2} combined with 
the previous expressions for the $x$-derivatives of $e_{\pm}$ 
gives $0=(u_{+}+ u_{-})\eta(e_{+},e_{-}) = -(u_{+}+ u_{-})$,
and hence we have $u_{+}=-u_{-}=u$.
The Frenet equations are thus given by 
\begin{equation}
\bmatr 
e_{+}\\e_{-}
\ematr_{x}
=\bmatr 
u&0\\0&-u
\ematr\bmatr 
e_{+}\\e_{-}
\ematr
\label{2dimnulleqs}
\end{equation}
where the Cartan matrix $\bmatr u&0\\ 0&-u \ematr$
belongs to the Lie algebra of the abelian group of null boosts in $\Mink{1}$.
These equations \eqref{2dimnulleqs} are preserved 
if the normalization \eqref{2dimnull2} of the null frame is changed. 
However, if the null parameter $x$ is changed by reparameterizing the curve, 
then the equations \eqref{2dimnulleqs} undergo a gauge transformation. 
In particular, for $x\rightarrow \tilde x(x)$, with $\tilde x'\neq 0$, 
we have $u\rightarrow \tilde u = (1/\tilde x') u + (1/\tilde x')'$. 
By restricting such reparameterizations to affine transformations 
$\tilde x=\alpha x +\beta$, where $\alpha\neq0$ and $\beta$ are constants,
we see that $u$ only undergoes a scaling by $\alpha$. 
Therefore, $u$ geometrically represents 
a relative affine invariant of the parameterized null curve.

We now consider {\em inelastic null} curve flows $\pos(t,x)$,
in which both the null signature $\eta(\pos_x,\pos_x)=0$ 
and the null parameter $x$
of the curve are locally preserved by the flow. 
Such flows are specified by a flow vector 
\begin{equation}\label{2dimnullflow}
\pos_t=h_{+} e_{+} + h_{-} e_{-}
\end{equation}
expressed in terms of the null frame \eqref{2dimnullframe}. 
The null frame will be carried by the flow, 
such that 
\begin{equation}
\bmatr e_{+}\\e_{-} \ematr_{t}
=\bmatr \omega& 0 \\ 0&-\omega \ematr\bmatr e_{+}\\e_{-} \ematr
\label{2dimnullfloweqs}
\end{equation}
where the Cartan matrix belongs to the Lie algebra of 
the abelian group of null boosts in $\Mink{1}$.
These equations \eqref{2dimnullfloweqs} 
arise from the null frame equations \eqref{2dimnull1}--\eqref{2dimnull2} 
similarly to the derivation of the Frenet equations \eqref{2dimnulleqs}
and are therefore preserved 
if the normalization \eqref{null2} of the null frame is changed. 
Consequently, $\omega$ represents 
a relative affine invariant of the time-parameterized curve flow.

The Frenet equations \eqref{2dimnulleqs} and the null flow equations \eqref{2dimnullfloweqs} 
are related by the compatibility condition that the mixed $t,x$-derivatives 
of the null frame \eqref{2dimnullframe} are equal. 
Likewise, 
the tangent vector \eqref{2dimnullT} and the flow vector \eqref{2dimnullflow}
of the curve are related by the compatibility condition that 
the mixed $t,x$-derivatives of $\pos(t,x)$ are equal. 
By the same steps used in the spacelike/timelike case, 
these conditions yield
\begin{gather}
u_t=\omega_x
\label{2dimnullstructeq1}
\\
\omega = h_{+}{}_x+uh_{+}
\label{2dimnullstructeq2}
\\
h_{-}{}_x = uh_{-}
\label{2dimnullstructeq3}
\end{gather}
which are the Cartan structure equations of the null frame. 

From equation \eqref{2dimnullstructeq3}, 
we naturally obtain a potential variable related to the variable $u$ by 
\begin{equation}\label{2dimnullv}
v_x = h_{-}{}_x/h_{-} = u ,
\quad
v= \ln h_{-} = \txtint udx
\end{equation}
up to the gauge freedom $v\rightarrow v+f(t)$. 
The remaining equations \eqref{2dimnullstructeq1} and \eqref{2dimnullstructeq2}
then become
\begin{equation} 
v_t=\omega = h_{+}{}_x+v_x h_{+} 
\label{2dimvfloweq}
\end{equation}
after a suitable gauge transformation is used to absorb an arbitrary integration constant 
(which is a function of $t$). 
Note, since $u$ has the geometrical meaning of 
a relative affine invariant of the curve 
under reparameterizations $x\rightarrow \alpha x+\beta$,
where $\alpha\neq0$ and $\beta$ are constants,
we see that $v=\int u dx$ is invariant up to an additive constant. 
Therefore $v$ represents an affine covariant of the curve. 

The system \eqref{2dimnullv}--\eqref{2dimvfloweq}
has an interesting operator structure, involving a recursion operator, 
as follows.

\begin{thm}
\label{2dimnullthm}
For inelastic null curve flows in $\Mink{1}$, 
the affine covariant $v=\int udx$ satisfies the equation
\begin{equation}
v_t=\Rop(h_{+}), 
\quad 
\Rop=D_x+v_x
\label{2dimnullR}
\end{equation}
in terms of the tangential component $h_{+}$ of the flow, 
where $\Rop$ is the hereditary recursion operator for Burgers' equation
in potential form. 
\end{thm}

This recursion operator $\Rop$ can be used to derive a hierarchy of flows 
starting from a root flow $v_t=v_x$ 
corresponding to the generator of $x$-translations on $v$. 
The following main result shows that 
the odd flows have a gradient-energy structure,
while the even flows have a Hamiltonian structure. 

\begin{thm}
\label{2dimvhierarchy}
There is a hierarchy of integrable flows on $v(t,x)$
given by
\begin{align}
v_t & 
= \Rop^n(v_x), 
\quad
n=1,2,\ldots 
\label{2dimvflows}
\\& 
=\begin{cases} 
-\exp(-2v) \delta \mk{H}^{(l)}/\delta v, 
\quad
l=(n+1)/2, &
n=1,3,\ldots
\\ 
\Hop(\delta \mathfrak{H}^{(l)}/\delta v), 
\quad
l=n/2, & 
n=2,4,\ldots
\end{cases} 
\label{2dimvgradHamilflows}
\end{align}
(called the {\em $+n$ flow})
in terms of energies/Hamiltonians $\mk{H}^{(n)}=\int H^{(n)} dx$, 
where 
\begin{equation}
\Hop=-\exp(-v)D_x\exp(-v)
\end{equation}
is a Hamiltonian operator, 
and where 
\begin{equation}
H^{(l)}=
\tfrac{1}{2} (-1)^{l-1} (D_x^{l}\exp(v))^2, 
\quad 
l=1,2,\ldots
\label{2dimvHam}
\end{equation}
are local densities. 
\end{thm}

The $+1$ flow in this hierarchy \eqref{2dimvflows} is explicitly given by 
\begin{equation}\label{potBurgers}
v_t=v_{xx}+v_x{}^2 
\end{equation}
which is Burgers' equation in potential form. 
It has the gradient structure
\begin{equation}
v_t=-\exp(-2v) \delta \mk{H}^{(1)}/\delta v 
\end{equation}
in terms of the energy density
\begin{equation}\label{vH1}
H^{(1)} = \tfrac{1}{2}\exp(2v)v_x{}^2 . 
\end{equation}
We remark that this structure implies 
\begin{equation}
\frac{d\mk{H}^{(1)}}{dt}
= \txtint v_t (\delta H^{(1)}/\delta v) dx
= -\txtint \exp(-2v) (\delta H^{(1)}/\delta v)^2 dx < 0
\end{equation}
(modulo boundary terms in the integral)
whereby the energy integral 
$\mk{H}^{(1)}=\int \tfrac{1}{2}\exp(2v)v_x{}^2 dx>0$ 
is a positive, decreasing function of $t$. 
(This property can be used to show that solutions 
$v(t,x)$ of equation \eqref{potBurgers} having finite energy
are dispersive.)

The odd part of the hierarchy \eqref{2dimvflows} 
has a gradient-energy structure similar to this $+1$ flow. 
In contrast, the even part of the hierarchy is quite different. 

The $+2$ flow is given by 
\begin{equation}\label{potAiry}
v_t=v_{xxx}+3v_{xx}v_x+ v_x{}^3
\end{equation}
which has the Hamiltonian structure
\begin{equation}
v_t=\Hop(\delta \mk{H}^{(1)}/\delta v)
\end{equation}
where the Hamiltonian is the same expression
$\mk{H}^{(1)}=\int \tfrac{1}{2}\exp(2v)v_x{}^2 dx$ 
as the energy integral appearing in the $+1$ flow. 
There is a similar Hamiltonian structure for the other even flows. 

The entire hierarchy \eqref{2dimvflows} of flows on $v(t,x)$ 
corresponds to a hierarchy of commuting vector fields 
\begin{equation}\label{2dimvsymms}
\X^{(n)} = \Rop^n(v_x)\partial_v,
\quad
n=0,1,2,\ldots
\end{equation}
with the root vector field being the generator of $x$-translations, 
$\X^{(0)} = v_x\partial_v$. 
From \thmref{2dimvhierarchy}, 
we see that all of the odd vector fields (\ie/ $n=1,3,\ldots$)
have a gradient structure, 
while all of the even vector fields (\ie/ $n=2,4,\ldots$)
have a Hamiltonian structure. 
Since the hierarchy is commuting, 
every vector field \eqref{2dimvsymms} in this hierarchy 
is the generator of a symmetry 
for Burgers' equation in potential form \eqref{potBurgers}. 

It is interesting to formulate the preceding results by using 
the variables $u$ and $h_{-}$ which are related to $v$ 
by equation \eqref{2dimnullv}. 
This equation has exactly the form of the Hopf-Cole transformation, 
under which Burgers' equation can be mapped to the heat equation. 

In terms of $u$ and $h_{-}$, 
the vector fields \eqref{2dimvsymms} become
\begin{equation}\label{2dimnullX}
\X^{(n)} = h_{+}^{(n)} \partial_v = g^{(n)} \partial_u = k^{(n)} \partial_{h_{-}},
\quad
n=0,1,2,\ldots
\end{equation}
where
\begin{equation}\label{2dimnullsymms}
h_{+}^{(n)} = \Rop^n(v_x), 
\quad
g^{(n)} =D_x h_{+}^{(n)} = \Qop^n(u_x), 
\quad
k^{(n)} = h_{-}h_{+}^{(n)} =\Sop^n(h_{-}{}_x)
\end{equation}
are given by the recursion operators
\begin{equation}\label{2dimnullRop}
\Rop = D_x + v_x,
\quad
\Qop = D_x\Rop D_x^{-1} = D_x+u+u_xD_x^{-1},
\quad
\Sop = \exp(v)\Rop\exp(-v) = D_x
\end{equation}
using equation \eqref{2dimnullv}. 
The odd and even vector fields in this hierarchy \eqref{2dimnullX}
will respectively inherit a gradient structure and a Hamiltonian structure 
through the variational relation
\begin{equation}
\delta\mk{H}/\delta v = -D_x\delta\mk{H}/\delta u = \exp(v)\delta\mk{H}/\delta h_{-} . 
\end{equation}
This leads to the following result. 

\begin{thm}
\label{2dimhuhierarchy}
The hierarchy of integrable flows \eqref{2dimvflows} on $v(t,x)$
yields the equivalent hierarchy of linear flows on $h_{-}(t,x)$, 
\begin{align}
h_{-}{}_t & 
= D_x{}^n(h_{-}{}_x), 
\quad
n=1,2,\ldots 
\label{2dimhflows}
\\& 
=\begin{cases} 
-\delta \mk{H}^{(l)}/\delta h_{-}, 
\quad
l=(n+1)/2, &
n=1,3,\ldots
\\ 
-D_x(\delta \mathfrak{H}^{(l)}/\delta h_{-}), 
\quad
l=n/2, & 
n=2,4,\ldots
\end{cases} 
\label{2dimhgradHamilflows}
\end{align}
in terms of energies/Hamiltonians $\mk{H}^{(l)}=\int H^{(l)} dx$, 
where
\begin{equation}
H^{(l)}=
\tfrac{1}{2} (-1)^{l-1} (D_x^{l} h_{-})^2
\quad 
l=1,2,\ldots
\label{2dimhHam}
\end{equation}
are local densities. 
Under the Hopf-Cole transformation \eqref{2dimnullv}, 
the hierarchy \eqref{2dimhflows} is mapped to 
\begin{align}
u_t & 
= \Qop^n(u_x), 
\quad
n=1,2,\ldots 
\label{2dimuflows}
\\& 
=\begin{cases} 
\Kop(\delta \mk{H}^{(l)}/\delta u), 
\quad
l=(n+1)/2, &
n=1,3,\ldots
\\ 
\Hop(\delta \mathfrak{H}^{(l)}/\delta u), 
\quad
l=n/2, & 
n=2,4,\ldots
\end{cases} 
\label{2dimugradHamilflows}
\end{align}
where $\mk{H}^{(l)}=\int H^{(l)} dx$ is given by the densities 
\begin{equation}
H^{(l)}=
\tfrac{1}{2} (-1)^{l-1} \exp(2v)((D_x+u)^{l-1} u)^2
\quad 
l=1,2,\ldots
\label{2dimuHam}
\end{equation}
and where 
\begin{equation}
\Kop =D_x\exp(-2v)D_x,
\quad
\Hop=\Qop\Kop= D_x\exp(-v)D_x\exp(-v)D_x
\end{equation}
are a positive symmetric operator and a Hamiltonian operator, 
respectively. 
\end{thm}

Corresponding to the potential form of Burgers' equation \eqref{potBurgers}, 
the $+1$ flow in these two hierarchies \eqref{2dimuflows} and \eqref{2dimhflows}
is given by the heat equation
\begin{equation}\label{heat}
h_{-}{}_t=h_{-}{}_{xx}
\end{equation}
and Burgers' equation (up to a scaling of $u$)
\begin{equation}\label{Burgers}
u_t=u_{xx}+2uu_x . 
\end{equation}
Both of these equations are integrable and have a gradient structure
\begin{equation}
h_{-}{}_t=-\delta \mk{H}^{(1)}/\delta h_{-}, 
\quad
u_t=\Kop(\delta \mk{H}^{(1)}/\delta u)
\end{equation}
in terms of the energy density
\begin{equation}\label{huH1}
H^{(1)} = \tfrac{1}{2} h_{-}{}_x{}^2 = \tfrac{1}{2}\exp(2v)u^2 . 
\end{equation}
All of the odd flows have a similar structure. 

The $+2$ flow in the hierarchies \eqref{2dimuflows} and \eqref{2dimhflows}
is given by 
\begin{equation}\label{AiryBurgers}
u_t=u_{xxx}+3(uu_{xx}+ u_x{}^2 + u^2u_x)
\end{equation}
and 
\begin{equation}\label{Airy}
h_{-}{}_t=h_{-}{}_{xxx}
\end{equation}
which is the Airy equation. 
These two integrable equations have the Hamiltonian structure
\begin{equation}
h_{-}{}_t=-D_x(\delta \mk{H}^{(1)}/\delta h_{-}), 
\quad
u_t=\Kop(\delta \mk{H}^{(1)}/\delta u)
\end{equation}
where the Hamiltonian is the same expression
as the energy integral appearing in the $+1$ flow. 
There is a similar structure for all of the even flows. 

\begin{prop}
Each flow in both hierarchies \eqref{2dimuflows} and \eqref{2dimhflows}
determines a null inelastic curve flow 
\begin{equation}\label{2dimnullcurveflow}
\pos_t=h_{+}^{(n)}e_{+} + h_{-}^{(n)}e_{-},
\quad
n=1,2,\ldots
\end{equation}
whose components are given by 
\begin{equation}\label{2dimnullcurveflowh}
\begin{aligned}
& 
h_{+}^{(n)} = h_{-}^{-1} D_x^n(h_{-}) = (D_x+u)^{n-1}u= \Rop^{n-1}(v_x),
\\
&
h_{-}^{(n)} = h_{-} = \exp(\txtint udx) = \exp(v),
\\
&
n=1,2,\ldots
\end{aligned}
\end{equation}
where 
\begin{equation}\label{2dimnullpair}
e_{+}=\pos_x,
\quad
e_{-}=\Nop(\pos_x) . 
\end{equation}
The expression 
\begin{equation}
u=\pm\eta(e_{\pm},e_{\mp}{}_x)=-\eta(\Nop(\pos_x),\pos_{xx})
\end{equation} 
shows that $u$ and its $x$-derivatives are invariant 
under translations and boosts applied to the curve $\pos$ 
in the Minkowski plane, 
so thus the equations of motion \eqref{2dimnullcurveflow} 
are invariant under the isometry group $ISO(1,1)$ of the Minkowski plane. 
\end{prop}

These equations of motion \eqref{2dimnullcurveflow} 
describe geometrical non-stretching motions of the curve. 
An interesting geometrical aspect of the motion emerges 
when the flow vector $\pos_t$ at each point $x$ on the curve 
is decomposed into complementary null components 
with respect to the light cone, 
\begin{equation}
(\pos_t)_{+}= ((D_x+u)^{n-1}u) e_{+} , 
\quad
n=1,2,\ldots
\label{2dimnullcurveflow+}
\end{equation}
and 
\begin{equation}
(\pos_t)_{-} = h_{-} e_{-} 
\label{2dimnullcurveflow-}
\end{equation}
where $(\pos_t)_{\pm}$ is a null vector parallel to $e_{\pm}$. 
Now, from the frame structure equations 
\eqref{2dimnulleqs}, 
\eqref{2dimnullfloweqs}, 
\eqref{2dimnullv}--\eqref{2dimvfloweq},
we have 
$(h_{-} e_{-})_x = (h_{-}{}_x-u h_{-}) e_{-} =0$ 
and 
$(h_{-} e_{-})_t = (h_{-}{}_t-\omega h_{-}) e_{-} =0$. 
Hence $h_{-} e_{-}$ is a constant null vector, 
which lies on the opposite side of the light cone relative to 
the tangent vector of the curve, $\pos_x$, 
as shown by equation \eqref{2dimnullpair}. 
This component \eqref{2dimnullcurveflow-} of the flow vector 
is the same for all of the flow equations \eqref{2dimnullcurveflow}. 
Moreover, going back to the original form of the frame structure equations
\eqref{2dimnullstructeq1}--\eqref{2dimnullstructeq3},
we see that this system has a consistent reduction if we put $h_{-}=0$,
or correspondingly, $(\pos_t)_{-}=0$. 
Under this reduction, 
the main results given by theorems~\ref{2dimnullthm} and~\ref{2dimvhierarchy}
continue to hold, as does the second half of \thmref{2dimhuhierarchy},
while only the Hopf-Cole transformation relating $u$ to $h_{-}$ is lost. 

The tangential component \eqref{2dimnullcurveflow+} of the flow vector 
determines a null inelastic curve flow by itself,
which we denote
\begin{equation}
(\pos_t)_\parallel= h_{+}^{(n)} T, 
\quad
n=1,2,\ldots
\label{2dimnulltangentialflow}
\end{equation}
in terms of the tangent vector of the curve
\begin{equation}
T = \pos_x
\end{equation} 
with 
\begin{equation}
h_{+}^{(n)} = (D_x+u)^{n-1}u . 
\end{equation} 
All of these curve flows 
preserve the null condition $\eta(\pos_x,\pos_x)=0$
and exhibit invariance under the isometry group $ISO(1,1)$. 
Thus, each equation of motion \eqref{2dimnulltangentialflow}
describes a geometric non-stretching motion of the curve. 

The $n=1$ curve flow \eqref{2dimnulltangentialflow} 
corresponding to the heat equation \eqref{heat}, 
and equivalently to Burgers' equation \eqref{Burgers} 
or its potential form \eqref{potBurgers}, 
is given by 
\begin{equation}\label{heatcurveflow}
(\pos_{t})_\parallel =u T .
\end{equation}
Through the frame structure equation \eqref{2dimnulleqs}, 
we can express this equation of motion \eqref{heatcurveflow}
in the simple form 
\begin{equation}
(\pos_{t})_\parallel =\pos_{xx}, 
\quad 
\eta(\pos_x,\pos_x)= 0
\end{equation} 
which is a non-stretching heat map equation in the Minkowski plane. 

Similarly, the $n=2$ curve flow \eqref{2dimnulltangentialflow} 
corresponding to the Airy equation \eqref{Airy}, 
and equivalently to equations \eqref{AiryBurgers} and \eqref{potAiry} 
obtained through the Hopf-Cole transformation, 
is given by 
\begin{equation}\label{airycurveflow}
(\pos_{t})_\parallel =(u_x+u^2) T .
\end{equation}
This equation of motion \eqref{airycurveflow}
can be expressed entirely in terms of $x$-derivatives of $\pos_x$
by use of the frame structure equation \eqref{2dimnulleqs}, as follows. 
We have $u_x \pos_x = (u\pos_x)_x-u\pos_{xx}= \pos_{xxx} - u^2\pos_x$,
which yields 
\begin{equation}
(\pos_{t})_\parallel =\pos_{xxx}, 
\quad 
\eta(\pos_x,\pos_x)= 0 . 
\end{equation} 
This curve flow is a non-stretching Airy map equation in the Minkowski plane.

\section{Timelike curve flows and integrable systems in $\Mink{2}$}
\label{3dimtimelike}

Three-dimensional Minkowski space is a vector space $\Mink{2}$
equipped with a Lorentz-signature $(-1,1,1)$ metric $\eta$
and a compatible volume form $\epsilon$.  
A vector $\vec v\in \Mink{2}$ is respectively 
{\em timelike}, {\em spacelike}, or {\em null} if 
its Minkowski norm $\eta(\vec v,\vec v)$ is negative, positive, or zero. 
The set of all null vectors spans a 2-dimensional null surface 
through the origin in $\Mink{2}$, called the {\em lightcone}. 

Up to a sign, the volume form is determined from the metric by the relation 
\begin{equation}
\begin{aligned}
\epsilon(\vec v,\vec w,\vec x)^2
= & \eta(\vec v,\vec w)^2\eta(\vec x,\vec x) 
+ \eta(\vec x,\vec v)^2\eta(\vec w,\vec w) 
+ \eta(\vec w,\vec x)^2\eta(\vec v,\vec v) 
\\&\qquad
- 2\eta(\vec v,\vec w)\eta(\vec w,\vec x)\eta(\vec x,\vec v)
- \eta(\vec v,\vec v)\eta(\vec w,\vec w)\eta(\vec x,\vec x)
\end{aligned}
\end{equation}
holding for any vectors $\vec v,\vec w,\vec x \in \Mink{2}$.
Associated to the metric and the volume form is 
the Hodge dual operator $*$ defined by 
\begin{equation}\label{3dimdual}
\eta(*(\vec v\wedge\vec w),\vec x) = \epsilon(\vec v,\vec w,\vec x) . 
\end{equation}
This operator maps pairs of vectors into vectors,
and thus it can be alternatively formulated as a Lorentzian version of 
the cross product defined by 
\begin{equation}\label{3dimcross}
\vec v\times\vec w = *(\vec v\wedge\vec w)
\end{equation}
which has the properties
\begin{equation}\label{cross1}
\vec v\times\vec v=0,
\quad
(\vec v\times\vec w)\times\vec x 
= \eta(\vec w,\vec x)\vec v - \eta(\vec v,\vec x)\vec w
\end{equation}
and 
\begin{equation}\label{cross2}
\eta(\vec v\times\vec w,\vec v) = \eta(\vec v\times\vec w,\vec w) = 0,
\quad
\eta(\vec v\times\vec w,\vec v\times\vec w) = 
\eta(\vec v,\vec w)^2-\eta(\vec v,\vec v)\eta(\vec w,\vec w)
\end{equation}
holding for any vectors $\vec v,\vec w,\vec x \in \Mink{2}$.

In inertial coordinates $ X^{i}=(X^{0},X^{1},X^{2})$, 
the Minkowski metric and the volume tensor respectively 
have the components
\begin{gather}
\eta_{ij}={\rm diag}(-1,1,1)
=\begin{cases} -1 & i=j=0\\ \hphantom{-}1 & i=j=1,2\\ \hphantom{-}0 & i\neq j\end{cases}
\\
\epsilon_{ijk}={\rm sgn}\bmatr i & j & k \\ 0 & 1 & 2\ematr
=\begin{cases} \hphantom{-}1 & (i,j,k)=(0,1,2), \text{cyclic}\\ -1 & (i,j,k)=(2,1,0), \text{cyclic}\\ \hphantom{-}0 & \text{otherwise} \end{cases}
\end{gather}
while the Hodge dual operator $*$ and the cross product $\times$
are both given by the associated tensor 
$\epsilon^i{}_{jk}=\epsilon_{ljk}\eta^{il}$.  
Here $X^0$ is a timelike coordinate and $X^1,X^2$ are spacelike coordinates. 

The isometries of 3-dimensional Minkowski space 
are given by the Poincare group 
$ISO(2,1)\simeq SO(2,1)\ltimes \Rnum^3$,
which comprises time and space translations, rotations, and boosts.

\subsection{Timelike curves and parallel frames}
\label{3dimtimelikeframe}
Let $\pos(x)=(X^0(x),X^1(x),X^2(x))$ be a timelike curve in $\Mink{2}$,
with the tangent vector $\pos_x$ satisfying 
\begin{equation}
\eta(\pos_x,\pos_x)<0
\label{timelikeT}
\end{equation}
at every point $x$ on the curve. 
Hereafter we choose $x$ to be the proper-time arclength parameter defined by 
\begin{equation}
|\pos_{x}|=\sqrt{-\eta(\pos_{x},\pos_{x})}=1 . 
\label{propertime}
\end{equation} 
Then 
\begin{equation}
T=\pos_x, \quad\text{unit timelike tangent vector}
\label{timeliketangent}
\end{equation}
satisfies
\begin{equation}
\eta(T,T)=-1
\label{timelikeunitT} . 
\end{equation}
Hence the normal plane orthogonal to $T$ at each point on the curve 
is a spacelike (Euclidean) plane, $\Rnum^2$. 

We begin by introducing the Lorentzian analog of a Frenet frame for such curves \cite{OzdErg}. 
The $x$-derivative of $T$ defines the principal normal vector $T_x$,
which is spacelike since it lies in the normal plane, 
due to $\eta(T,T_x)=0$. 
Therefore
\begin{equation}
N=\kappa^{-1}T_x, \quad\text{unit spacelike normal vector}
\label{timelikenormal}
\end{equation}
satisfies
\begin{equation}
\eta(T,N)=0,
\quad
\eta(N,N)=1
\label{timelikeunitN}
\end{equation}
where the function $\kappa(x)$ is given by 
\begin{equation}
\kappa=\eta(T_x,N)= \sqrt{\eta(T_x,T_x)}
\label{timelikecurvinv}
\end{equation}
which is the Lorentzian curvature invariant of the curve. 
The cross product of $T$ and $T_x$ yields a vector 
which is orthogonal to both $T$ and $T_x$ from property \eqref{cross2}. 
Since this vector $T\times T_x$ lies in the normal plane,
it is spacelike.
Hence 
\begin{equation}
B=\kappa^{-1} T\times T_x, \quad\text{unit spacelike bi-normal vector}
\label{timelikebinormal}
\end{equation}
satisfies
\begin{equation}
\eta(B,T)=\eta(B,N)=0,
\quad
\eta(B,B)=1
\label{timelikeunitB}
\end{equation}
The triple of mutually orthogonal vectors 
\begin{equation}
\bmatr 
T\\  N \\ B
\ematr
\label{timelikefrenetTNB}
\end{equation}
defines a Frenet frame for a timelike curve $\pos(x)$ in $\Mink{2}$. 
Note that the orientation of this frame is 
\begin{equation}\label{timelikefrenetoriented}
\epsilon(T,N,B) = 1
\end{equation}
in terms of the volume form. 

The Frenet equations of this frame \eqref{timelikefrenetTNB} 
are straightforward to derive. 
First, from equation \eqref{timelikenormal}, we have
\begin{equation}
T_x=\kappa N .
\label{timelikeTx}
\end{equation}
Next, 
the $x$-derivative of equation \eqref{timelikeunitN} yields
$0=\eta(N_x,N)$ and $0=\eta(T_x,N)+\eta(T,N_x)= \kappa + \eta(N_x,T)$,
so thus we have 
\begin{equation}
N_x=\kappa T+\tau B
\label{timelikeNx}
\end{equation}
for some function $\tau(x)$. 
To obtain an expression for $\tau$ in terms of $T$ and its $x$-derivatives, 
we substitute equations \eqref{timelikenormal} and \eqref{timelikebinormal}
into equation \eqref{timelikeNx} 
and take its Minkowski inner product with $B$. 
This yields
\begin{equation}
\tau=\eta(N_x,B)=\kappa^{-2}\eta(T\times T_x,T_{xx})
\label{timeliketorsinv}
\end{equation}
which is the Lorentzian torsion invariant of the curve. 
Last, from the $x$-derivative of equation \eqref{timelikebinormal}, 
we find $B_x=T_x\times N+T\times N_x= \tau T\times B$
after using equations \eqref{timelikeTx} and \eqref{timelikeNx}. 
Then using property \eqref{cross1}, we have
\begin{equation}
B_x=-\tau N .
\label{timelikeBx}
\end{equation} 

Therefore, the Frenet equations are given by
\begin{equation}
\bmatr 
T_x\\  N_x \\ B_x
\ematr=\bmatr 
0 & \kappa & 0\\ \kappa & 0 & \tau \\ 0 & -\tau
& 0
\ematr\bmatr 
T\\  N \\ B
\ematr
\label{timelikefreneteqns}
\end{equation}
where $\bmatr 0 & \kappa & 0\\ \kappa & 0 & \tau \\ 0 & -\tau & 0 \ematr\in \mathfrak{so}(2,1)$ is the Cartan matrix,
which belongs to the Lie algebra $\mathfrak{so}(2,1)$ of the 
$SO(2,1)$ group of rotation and boost isometries in $\Mink{2}$.

A general frame for a timelike curve in $\Mink{2}$ is related to 
this Frenet frame by the action of arbitrary $x$-dependent $SO(2,1)$ 
rotations and boosts applied to the vectors \eqref{timelikefrenetTNB}. 
If the tangent vector $T$ is preserved as one of the frame vectors, 
then the resulting frame is given by 
applying a general $x$-dependent $SO(2)$ rotation 
on the normal vectors $N$ and $B$ in the spacelike normal plane of the curve. 
This yields
\begin{equation}
\E=\bmatr {e}_0\\{e}_1 \\{e}_2 \ematr
=
\bmatr 
1&0           &0            \\
0&\cos \theta &\sin \theta  \\
0&-\sin \theta &\cos \theta\\
\ematr
\bmatr T\\N\\B\\\ematr
=\bmatr 
T \\ (\cos\theta) N+(\sin\theta) B \\ -(\sin\theta) N+(\cos\theta)B
\ematr
\label{timelikeadaptedframe}
\end{equation}
where $\theta(x)$ is the rotation angle,
and where the frame vectors satisfy the orthogonality relations
\begin{gather}
\eta(e_0,e_0)=-1,
\quad 
\eta(e_1,e_1)=\eta(e_2,e_2)=1,
\label{timelike1}\\
\eta(e_0,e_1)=\eta(e_0,e_2)=\eta(e_1,e_2)=0
\label{timelike2}
\end{gather}
and the cross product relations
\begin{equation}
e_1\times e_2=-e_0,
\quad
e_0\times e_1=e_2,
\quad
e_0\times e_2=-e_1 
\end{equation}
which follow from $\epsilon(e_0,e_1,e_2)=\epsilon(T,N,B) = 1$. 

We now use this gauge freedom to define a {\em Lorentzian parallel frame}
by the geometrical condition that the $x$-derivative of 
both normal vectors $e_1$ and $e_2$ is parallel to the tangent vector $e_0$. 
Thus we require 
 \begin{equation}
e_1{}_x=u_1e_0, 
\quad 
e_2{}_x=u_2e_0
\label{timelikeparallel1}
\end{equation}
for some functions $u_1(x),u_2(x)$. 
This determines part of the Cartan matrix. 
To derive the remaining part, we need to work out $e_0{}_x$. 
By substituting equation \eqref{timelikeparallel1}
into the $x$-derivative of equations \eqref{timelike1} and \eqref{timelike2},
we get 
$\eta(e_0{}_x,e_1)= u_1$, 
$\eta(e_0{}_x,e_2)=u_2$, 
and $\eta(e_0{}_x,e_0)=0$. 
These relations determine 
\begin{equation}
e_0{}_x=u_1e_1+u_2e_2 . 
\label{timelikeparallel2}
\end{equation}
Therefore, the Cartan matrix of a Lorentzian parallel frame 
for a timelike curve in $\Mink{2}$ is given by 
\begin{equation}
\E_x=\U\E
\label{timelikeEx}
\end{equation}
where 
\begin{equation}
\U=\bmatr 
0 & u_1& u_2\\ u_1 & 0 & 0\\ u_2 & 0 & 0
\ematr
\in \mk{so}(2,1) . 
\label{timelikeparallelC}
\end{equation}
This matrix belongs to the perp space of the stabilizer subalgebra 
$\mk{so}(2)\subset \mk{so}(2,1)$ of the frame vector $e_0$. 
In particular, there is a decomposition 
$\mk{so}(2,1) = \mk{so}(2)\oplus \mk{so}(2)_\perp$ 
as a symmetric Lie algebra,
where $\mk{so}(2)$ is the subalgebra of rotations 
and $\mk{so}(2)_\perp\simeq \Rnum^2$ is its perp space given by 
the span of the 1-dimensional subalgebras of boosts 
which act in two orthogonal planes containing $e_0$.

To determine the rotation angle $\theta(x)$ 
under which a Frenet frame is transformed to a parallel frame, 
we first need the inverse transformation
\begin{equation}
\bmatr T\\N \\B\ematr
=
\bmatr 
e_0 \\
(\cos\theta) e_1 -(\sin\theta)e_2  \\
(\sin\theta) e_1+(\cos\theta) e_2 \\
\ematr . 
\label{timelikeinvtrans}
\end{equation}
Now, we take the $x$-derivative of the frame \eqref{timelikeadaptedframe} 
and substitute the Frenet equations \eqref{timelikefreneteqns} 
followed by the inverse transformation \eqref{timelikeinvtrans}, 
which yields
\begin{equation}
\E_x=
\bmatr 
(\kappa \cos\theta) e_1 -(\kappa\sin\theta) e_2\\
(\kappa \cos\theta) e_0+(\theta_x+\tau) e_2  \\
-(\kappa \sin\theta) e_0-(\theta_x+\tau) e_1 
\ematr . 
\end{equation}
Thus the condition \eqref{timelikeEx}--\eqref{timelikeparallelC}
can be achieved if (and only if)
\begin{equation}
\theta_x = -\tau . 
\label{timelikerotationcond}
\end{equation} 
The resulting frame given by equations \eqref{timelikeadaptedframe} 
and \eqref{timelikerotationcond} thereby defines a parallel frame, 
where the components of its Cartan matrix are related to 
the curvature and torsion invariants $\kappa,\tau$ by 
\begin{equation}
u_1=\kappa\cos\theta=\kappa \cos\left(-\txtint \tau dx\right),
\quad 
u_2=-\kappa\sin\theta=\kappa \sin\left(\txtint \tau dx\right) . 
\end{equation}
These expressions are a Lorentzian counterpart of the well-known 
Hasimoto transformation in Euclidean space. 

A parallel frame is unique up to $x$-independent (rigid) rotations 
\begin{equation}
\theta\rightarrow\theta+\phi
\end{equation}
where $\phi$ is constant.
Under these rotations, 
the tangent vector $e_0$ is preserved,
while the normal vectors $e_1$ and $e_2$ are rigidly rotated
\begin{equation}
\begin{aligned}
& e_1\rightarrow (\cos\phi)e_1+(\sin\phi)e_2,
\\
& e_2\rightarrow -(\sin\phi)e_1+(\cos\phi)e_2 . 
\end{aligned}
\label{timelikerigidrotation}
\end{equation}
Any two parallel frames related by this transformation are gauge equivalent. 
The $SO(2)$ group of rigid rotations thereby defines 
the gauge (equivalence) group for parallel frames. 
Stated another way, a parallel frame is determined by a timelike curve 
only up to the action of this gauge group.

\subsection{Inelastic Flow Equations}
We now consider curve flows $\pos(t,x)$ that locally preserve 
both the timelike signature \eqref{timelikeT} of the curve 
and the proper time normalization \eqref{propertime} of the arclength parameter.
Such flows are called {\em inelastic}
and are specified by a flow vector 
\begin{equation}\label{timelikeflow}
\pos_t=h_\parallel e_0+h_1e_1+h_2e_2
\end{equation}
expressed in terms of a tangential component $h_\parallel$ 
and a pair of normal components $h_1,h_2$ 
with respect to the frame vectors $e_0$, $e_1$, $e_2$. 

The parallel frame will be carried by the flow, 
such that the orthogonality relations \eqref{timelike1}--\eqref{timelike2}
are preserved. 
This implies that the $t$-derivative of the frame vectors 
$e_0$, $e_1$, $e_2$ is given by 
\begin{gather}
e_0{}_t=\omega_1e_1+\omega_2e_2,
\\
e_1{}_t=\omega_1e_0+\omega_0e_2,
\quad 
e_2{}_t=\omega_2e_0-\omega_0e_1.
\end{gather}
We can write these equations in the form
\begin{equation}
\E_t=\W \E
\label{timelikeEt}
\end{equation}
with the Cartan matrix
\begin{equation}
\W=\bmatr 
0 & \omega_1& \omega_2\\ \omega_1 & 0 & \omega_0\\ \omega_2 & -\omega_0 & 0
\ematr
\in \mk{so}(2,1)
\label{timelikeflowC}
\end{equation}
which belongs to the Lie algebra of the $SO(2,1)$ group of 
rotation and boost isometries in $\Mink{2}$.

The flow equations \eqref{timelikeEt} 
and the Frenet equations \eqref{timelikeEx}
of the parallel frame are related by the compatibility condition 
$\partial_t(\E_x)=\partial_x(\E_t)$. 
This condition is equivalent to a zero curvature equation
\begin{equation}
\U_t-\W_x+[\mathbf U,\mathbf W]=0
\label{timelikezerocurv}
\end{equation}
relating the Cartan matrices $\W$ and $\U$. 
After substituting these matrices \eqref{timelikeparallelC} and \eqref{timelikeflowC} 
into equation \eqref{timelikezerocurv}, 
we obtain
\begin{gather}
u_{1t}=\omega_{1x}+u_2\omega_0, 
\quad 
u_{2t}=\omega_{2x}-u_1\omega_0,
\label{timelikezerocurv1}\\
\omega_{0x}=u_1\omega_2-\omega_1u_2 . 
\label{timelikezerocurv2}
\end{gather}

Likewise, 
the flow vector \eqref{timelikeflow} 
and the tangent vector \eqref{timeliketangent} 
of the curve are related by the compatibility condition 
$\partial_x(\pos_t)=\partial_t(\pos_x)$. 
By writing 
\begin{equation}
\H=\bmatr h_\parallel \\ h_1 \\ h_2 \ematr
\in \Rnum^3,
\quad
\e= \bmatr 1 \\ 0 \\ 0 \ematr
\in \Rnum^3 , 
\label{timelikeHe}
\end{equation}
we have 
$\pos_x=\e^\t\E$
and 
$\pos_t=\H^\t\E$. 
Then the compatibility condition becomes 
\begin{equation}
\H_x+\U^\t\H = \W^\t\e
\label{timelikezerotors}
\end{equation}
relating $\W$ to $\U$ and $\H$. 
After we substitute the matrices \eqref{timelikeparallelC} and \eqref{timelikeflowC} 
along with the vectors \eqref{timelikeHe}
into equation \eqref{timelikezerotors}, 
we find that its tangential and normal components yield 
\begin{gather}
h_{\parallel}{}_x=-h_1u_1-h_2u_2, 
\label{timelikezerotors1}\\
\omega_1=h_1{}_x+u_1h_\parallel,
\quad 
\omega_2=h_2{}_x+u_2h_\parallel .
\label{timelikezerotors2}
\end{gather} 

These compatibility equations 
\eqref{timelikezerocurv1}, \eqref{timelikezerocurv2}, 
\eqref{timelikezerotors1}, \eqref{timelikezerotors2} 
are the Cartan structure equations of the parallel frame. 
They describe all inelastic timelike curve flows $\pos(t,x)$ in $\Mink{2}$.

\subsection{U(1)-invariant formalism}
\label{SO(2)invariance}
The gauge group for parallel frames consists of 
rigid $SO(2)$ rotations \eqref{timelikerigidrotation}
acting on the normal vectors. 
Under the action of this group, 
the components of the Cartan matrix \eqref{timelikeparallelC} 
along the curve are transformed by 
\begin{equation}
\begin{aligned}
& u_1\rightarrow u_1\cos\phi+u_2\sin\phi=\kappa \cos(\theta+\phi), 
\\
& u_2\rightarrow-u_1\sin\phi+u_2\cos\phi=-\kappa \sin(\theta+\phi) .
\end{aligned}
\label{timelikeuequiv}
\end{equation}
Similarly, the components of the Cartan matrix \eqref{timelikeflowC} 
along the flow are transformed by
\begin{gather}
\omega_0\rightarrow\omega_0,
\label{timelikew0equiv}
\\
\begin{aligned}
& \omega_1\rightarrow\omega_1\cos\phi+\omega_2\sin\phi,
\\
& \omega_2\rightarrow-\omega_1\sin\phi+\omega_2\cos\phi .
\end{aligned}
\label{timelikewequiv}
\end{gather}
Since the flow vector \eqref{timelikeflow} is gauge invariant, 
its normal and tangential components in a parallel frame 
are transformed by 
\begin{gather} 
h_\parallel\rightarrow h_\parallel,
\label{timelikehparequiv}
\\
\begin{aligned}
& h_1\rightarrow h_1\cos\phi+h_2\sin\phi,
\\
& h_2\rightarrow -h_1\sin\phi+h_2\cos\phi.
\end{aligned}
\label{timelikehperpequiv}
\end{gather}
Hence it is natural to introduce a complex formalism  
\begin{equation}
u=u_1+iu_2=\kappa e^{-i\theta}=\kappa \exp\left(i\txtint\tau dx\right),
\quad
\omega=\omega_1+i\omega_2,  \quad 
h_\perp=h_1+ih_2 
\end{equation}
in which the rigid rotations 
\eqref{timelikeuequiv}, \eqref{timelikewequiv}, \eqref{timelikehperpequiv}
become phase rotations
\begin{equation}
u\rightarrow u\exp(-i\phi),
\quad
\omega\rightarrow \omega\exp(-i\phi),
\quad 
h_\perp\rightarrow h_\perp\exp(-i\phi).
\end{equation}

Now, we can write the system 
\eqref{timelikezerocurv1}, \eqref{timelikezerocurv2}, 
\eqref{timelikezerotors1}, \eqref{timelikezerotors2} 
in a $U(1)$-invariant form:
\begin{gather}
u_t=\omega_x-iu\omega_0, 
\label{timelikeu}\\
\omega_{0x}=\Im({\bar u}\omega), 
\label{timelikew0}\\
h_\parallel{}_x=-\Re({\bar u }h_\perp), 
\label{timelikehpar}\\
\omega=h_\perp{}_x+uh_\parallel . 
\label{timelikew}
\end{gather}
Since $u$ is determined by the curve 
only up to rigid $U(1)$ phase rotations \eqref{timelikeuequiv}, 
this complex variable $u$ is a Hasimoto variable 
which geometrically represents a {\em $U(1)$-covariant} of the curve,
in contrast to the invariants $\kappa,\tau$
which are determined uniquely by the curve. 

This system \eqref{timelikeu}--\eqref{timelikew} has the following 
operator formulation which encodes a triple of $U(1)$-invariant Hamiltonian operators. 

\begin{thm}
\label{timelikeHJthm}
For timelike inelastic curve flows in $\Mink{2}$, 
the curve covariant $u$ satisfies the $U(1)$-invariant system 
\begin{align}
& u_t=D_x\omega-iu D_x^{-1}\Im(\omega \bar u) = \Hop(\omega)
\label{timelikeueqn}
\\
& \omega=D_{x}h_\perp-uD_{x}^{-1}\Re(\bar u h_\perp)=\Jop(h_\perp)
\label{timelikeweqn}
\end{align}
where 
\begin{align}
& \Hop=D_x +iuD_{x}^{-1}\Im(u\Cop)
\label{timelikeH}\\
& \Jop=D_{x} -uD_{x}^{-1}\Re(u\Cop)
\label{timelikeJ}
\end{align}
are a pair of compatible, $U(1)$-invariant
Hamiltonian and symplectic operators,
and $\Cop$ denotes the complex conjugation operator. 
Moreover, the operators $\Hop$ and $\Jop$ are related by 
\begin{equation}
\Hop\Iop^{-1}= -\Iop\Jop,
\quad
\Jop\Iop= -\Iop^{-1}\Hop,
\quad
\Iop =-i 
\label{timelikeI}
\end{equation}
where $\Iop$ is a Hamiltonian operator 
compatible with $\Hop$ and $\Jop$. 
Composition of these operators yields the $U(1)$-invariant evolution equation 
\begin{equation}\label{timelikeuflow}
u_t=-\Rop^2(h_\perp)
\end{equation}
for $u$ in terms of the normal component $h_\perp$ of the flow, 
where 
\begin{equation}
\Rop=\Hop\Iop^{-1} =-\Iop\Jop = i(D_x - uD_x^{-1}\Re(u\Cop))
\end{equation}
is a hereditary recursion operator. 
\end{thm}

This theorem is a counterpart of a similar result \cite{MarSanWan,AncMyr} 
for inelastic curve flows in $\Rnum^3$ 
and can be proved by the same methods. 
In particular, compared to the Euclidean case, 
there is only a change in the sign of the nonlocal term 
in both operators $\Hop$ and $\Jop$. 

An operator $\Dop$ here is Hamiltonian iff 
it defines an associated Poisson bracket
\begin{equation}\label{poisson}
\{\mk{H},\mk{G} \}_{\Dop} =
\int \Re\big( (\delta \mk{H}/\delta u) \Dop( \delta \mk{G}/\delta\bar u) \big) dx
\end{equation}
obeying skew-symmetry 
$\{\mk{H},\mk{G} \}_{\Dop} + \{\mk{G},\mk{H} \}_{\Dop} =0$
and the Jacobi identity 
$\{\{\mk{H},\mk{G} \}_{\Dop},\mk{F}\}_{\Dop} + \text{cyclic } =0$
for all functionals $\mk{H}$, $\mk{G}$, $\mk{F}$ 
on the $x$-jet space $J^\infty$ of the variables $u$ and $\bar u$, 
where $\delta /\delta u$ and $\delta /\delta\bar u$ 
denote the standard Euler operators. 
The formal inverse of a Hamiltonian operator defines a symplectic operator. 
Compatibility of two Hamiltonian operators $\Dop_1$ and $\Dop_2$
is the statement that every linear combination 
$c_1\Dop_1+c_2\Dop_2$ is a Hamiltonian operator. 

Due to a general theorem of Magri \cite{Mag}, 
the $U(1)$ invariance of the pair of compatible Hamiltonian operators 
$\Hop$ and $\Iop$
can be used to derive a hierarchy of flows starting from a root flow 
$u_t=-iu$ corresponding to the generator of $U(1)$ phase-rotations on $u$. 
This leads to the following main result. 

\begin{thm}
\label{timelikehierarchy}
There is a hierarchy of integrable tri-Hamiltonian flows on $u(t,x)$
given by
\begin{equation}
\begin{aligned}
u_t=\Rop^{n+1}(-iu) & 
= \Iop(\delta \mk{H}^{(n+1)}/\delta\bar u) 
=\Hop(\delta \mk{H}^{(n)}/\delta\bar u) , 
\quad 
n=0,1,2,\ldots 
\\& 
=\Eop(\delta \mathfrak{H}^{(n-1)}/\delta\bar u), 
\quad 
n=1,2,\ldots 
\end{aligned}
\label{timeliketrihamil}
\end{equation}
(called the {\em $+n$ flow})
in terms of Hamiltonians $\mk{H}^{(n)}=\int H^{(n)} dx$
where 
\begin{equation}
\begin{aligned}
& \Iop =-i,
\quad
\Hop=D_x +iuD_x^{-1}\Im(u\Cop),
\\
& \Eop = \Rop\Hop= i(D_x{}^2 -|u|^2 +uD_x^{-1}\Re(u_x\Cop)+iu_xD_x^{-1}\Im(u\Cop))
\end{aligned}
\end{equation}
are compatible Hamiltonian operators, 
and where 
\begin{equation}
H^{(n)}=
2(1+n)^{-1} D^{-1}_x\Im(\bar u(i\Hop)^{n+1} u), 
\quad 
n=0,1,2,\ldots
\label{timelikeHam}
\end{equation}
are local Hamiltonian densities. 
\end{thm}

The expression for the Hamiltonians \eqref{timelikeHam} in this theorem 
arises from a scaling formula derived in \Ref{Anc03}. 

The $+1$ flow in the hierarchy \eqref{timeliketrihamil} is explicitly given by 
\begin{equation}\label{timelikenls}
u_t=i(u_{xx}-\tfrac{1}{2}|u|^{2}u)
\end{equation}
which is the defocusing NLS equation.
It has the explicit tri-Hamiltonian structure
\begin{equation}\label{timelikenlsnlshamil}
u_t = \Iop(\delta\mk{H}^{(2)}/\delta\bar u) 
=\Hop(\delta \mk{H}^{(1)}/\delta\bar u) 
=\Eop(\delta\mk{H}^{(0)}/\delta\bar u)  
\end{equation}
in terms of the Hamiltonian densities 
\begin{equation}\label{timelikeH012}
H^{(0)} = |u|^2,
\quad
H^{(1)} = \Im(u\bar u_x),
\quad
H^{(2)} = |u_x|^2 + \tfrac{1}{4} |u|^4
\end{equation}
(modulo trivial densities given by total $x$-derivatives). 
Similarly, the $+2$ flow in the hierarchy \eqref{timeliketrihamil} is given by 
\begin{equation}\label{timelikemkdv}
u_t=u_{xxx}-\tfrac{3}{2}|u|^{2}u_x
\end{equation}
which is the complex defocusing mKdV equation. 
It has the tri-Hamiltonian structure
\begin{equation}\label{timelikemkdvhamil}
u_t = \Iop(\delta\mk{H}^{(3)}/\delta\bar u) 
=\Hop(\delta \mk{H}^{(2)}/\delta\bar u) 
=\Eop(\delta\mk{H}^{(1)}/\delta\bar u)  
\end{equation}
where $\mk{H}^{(1)}$ and $\mk{H}^{(2)}$ are given by 
the Hamiltonian densities \eqref{timelikeH012}, 
while $\mk{H}^{(3)}$ is given by the Hamiltonian density 
\begin{equation}\label{timelikeH3}
H^{(3)} = \Im(u_x\bar u_{xx}) +\tfrac{3}{4}|u|^2\Im(u\bar u_x)
\end{equation}
(modulo trivial densities given by total $x$-derivatives). 

The entire hierarchy \eqref{timeliketrihamil} of flows corresponds to 
a hierarchy of commuting vector fields 
\begin{equation}\label{timelikesymms}
\X^{(n)} = \Rop^n(-iu)\partial_u,
\quad
n=0,1,2,\ldots
\end{equation}
with the root vector field being the generator of phase rotations
$\X^{(0)} = -iu\partial_u$. 
In this setting, 
a vector field is Hamiltonian iff 
its prolongation on the $x$-jet space $J^\infty$ of $u$ and $\bar u$ satisfies
$\pr\X\mk{G} = \{\mk{G},\mk{H} \}_{\Dop}$ for some functional $\mk{H}$,
where $\mk{G}$ is an arbitrary functional,
and $\Dop$ is a given Hamiltonian operator. 
As shown by \thmref{2dimhierarchy}, 
all of the vector fields in this hierarchy are bi-Hamiltonian, 
with 
$\pr\X^{(n)}\mk{G} = \{\mk{G},\mk{H}^{(n)}\}_{\Hop}= \{\mk{G},\mk{H}^{(n+1)}\}_{\Iop}$
for $n\geq0$, 
where $\mk{H}^{(n)}=\int H^{(n)}dx$ is the functional 
with the Hamiltonian density \eqref{timelikeHam},
while each vector field for $n\geq 1$ is also tri-Hamiltonian, 
$\pr\X^{(n)}\mk{G} = \{\mk{G},\mk{H}^{(n-1)}\}_{\Eop}$. 
Since the entire hierarchy is commuting, 
every Hamiltonian vector field \eqref{timelikesymms} 
is the generator of a symmetry 
for the defocusing NLS equation \eqref{timelikenls}
as well as the defocusing mKdV equation \eqref{timelikemkdv}, 
and every associated Hamiltonian \eqref{timelikeHam} 
is a conserved density 
for both of these equations. 

\begin{prop}
Each tri-Hamiltonian flow in the hierarchy \eqref{timeliketrihamil} 
can be written in the form \eqref{timelikeuflow} as given by 
\begin{equation}
h_\perp=-\Rop^{n-1}(-iu), 
\quad 
n=1,2,\ldots . 
\end{equation}
Consequently, 
the $+n$ flow for $n\geq1$ determines an inelastic timelike curve flow 
\begin{equation}\label{timelikecurveflow}
\pos_t=h_\parallel^{(n)}e_0 + h_1^{(n)}e_1 + h_2^{(n)}e_2
\quad
n=1,2,\ldots
\end{equation}
whose tangential and normal components are given by 
\begin{equation}\label{timelikeflowh}
h_\parallel^{(n)}= \tfrac{1}{2}(n-1) H^{(n-2)},
\quad
h_1^{(n)}+ ih_2^{(n)} = -\Rop^{n-1}(-iu), 
\quad
n=1,2,\ldots . 
\end{equation}
\end{prop}

To write these equation of motions \eqref{timelikecurveflow} 
in an explicit form, 
it is convenient to introduce a complex parallel-frame notation:
\begin{equation}
e_\parallel=e_0 =T, 
\quad 
e_\perp=e_1+ie_2 = e^{-i\theta}(N+iB)
\label{timelikecomplexE}
\end{equation}
satisfying
\begin{equation}
e_\parallel\times e_\perp= -ie_\perp, 
\quad
e_\perp\times \bar e_\perp= i2e_\parallel
\label{timelikecrossE}
\end{equation}
and 
\begin{equation}
\eta(e_\parallel, e_\parallel) =-1, 
\quad
\eta(e_\perp,\bar e_\perp)= 2, 
\quad
\eta(e_\parallel,e_\perp)=\eta(e_\perp,e_\perp)=0 . 
\label{timelikecomplexEnorms}
\end{equation}
The action of the gauge group \eqref{timelikerigidrotation}
on these frame vectors consists of rigid $U(1)$ phase rotations
\begin{equation}\label{timelikecomplexequiv}
e_\parallel \rightarrow e_\parallel,
\quad 
e_\perp \rightarrow e^{-i\phi}e_\perp
\end{equation}
where $\phi$ is constant. 

The Frenet equations \eqref{timelikeEx} of the underlying parallel frame 
become
\begin{equation}
e_\parallel{}_x=\Re(\bar u e_\perp) ,\quad
e_\perp{}_x=ue_\parallel , 
\label{timelikecomplexEx}
\end{equation}
while the flow equations \eqref{timelikeEt} are similarly given by 
\begin{equation}
e_\parallel{}_t=\Re(\bar\omega e_\perp) ,
\quad
e_\perp{}_t=\omega e_\parallel  -i\omega_0 e_\perp 
\label{timelikecomplexEt}
\end{equation}

The equation of motions \eqref{timelikecurveflow} for the curve 
now take the form 
\begin{equation}\label{timelikecomplexcurveflow}
\pos_t=h_\parallel^{(n)}e_\parallel + \Re(\bar h_\perp^{(n)} e_\perp), 
\quad
n=1,2,\ldots
\end{equation}
where the tangential component $h_\parallel^{(n)}$
and the normal component $h_\perp^{(n)}=h_1^{(n)}+ ih_2^{(n)}$
are functions of the curve covariant $u$, 
the complex conjugate covariant $\bar u$,
and their $x$-derivatives, 
as given by the expressions \eqref{timelikeflowh}. 
Since $u\rightarrow e^{-i\phi}u$ and $\bar u\rightarrow e^{i\phi}\bar u$
under the gauge group \eqref{timelikecomplexequiv}, 
the tangential component is gauge invariant 
while the normal component is gauge equivariant, 
\begin{equation}
h_\parallel^{(n)} \rightarrow h_\parallel^{(n)},
\quad
h_\perp^{(n)} \rightarrow e^{-i\phi} h_\perp^{(n)} . 
\end{equation}
Consequently, 
each equation of motion \eqref{timelikeflow} is invariant under 
the isometry group $ISO(2,1)$ of $\Mink{2}$
and thus describes a geometric non-stretching motion of the curve.

The curve flow $n=1$ 
corresponding to the defocusing NLS equation \eqref{timelikenls}
is determined by the components
\begin{equation}
h_\parallel^{(1)}=0,
\quad
h_\perp^{(1)}=iu . 
\end{equation}
Substituting these expressions into the $n=1$ equation of motion \eqref{timelikeflow}, 
we have 
$\pos_t = \Re(-i\bar u e_\perp)
=\Re(\bar u e_\parallel\times e_\perp)
= e_\parallel\times e_\parallel{}_x$
by using the cross product identity \eqref{timelikecrossE} 
and the Frenet equations \eqref{timelikecomplexEx}. 
Thus we obtain 
\begin{equation}\label{timelikenlscurveflow}
\pos_{t}=T\times T_x = \pos_x\times\pos_{xx},
\quad 
\eta(\pos_x,\pos_x)=-1
\end{equation}
which is a timelike version of the vortex filament equation in 
Minkowski space $\Mink{2}$. 

The curve flow $n=2$ 
corresponding to the defocusing mKdV equation \eqref{timelikemkdv}
has the components
\begin{equation}
h_\parallel^{(2)}=\tfrac{1}{2}|u|^2,
\quad
h_\perp^{(2)}=-u_x . 
\end{equation}
Substituting these expressions into the $n=2$ equation of motion \eqref{timelikeflow}, 
we find 
$\pos_t = \Re(-\bar u_x e_\perp) +\tfrac{1}{2}|u|^2 e_\parallel$.
We simplify the term 
$\Re(-\bar u_x e_\perp)= -\Re(\bar u e_\perp)_x + \Re(\bar u e_\perp{}_x)
= -e_\parallel{}_{xx} + |u|^2 e_\parallel$
by using the Frenet equations \eqref{timelikecomplexEx}. 
Next we note 
$\eta(e_\parallel{}_x,e_\parallel{}_x) 
=\eta(\Re(\bar u e_\perp),\Re(\bar u e_\perp))
= \tfrac{1}{2} \eta(\bar u e_\perp,u\bar e_\perp)= |u|^2$
which follows from the relations \eqref{timelikecomplexEnorms}. 
This yields 
$\pos_t = -e_\parallel{}_{xx} +\tfrac{3}{2}|u|^2 e_\parallel$,
and thus we obtain 
\begin{equation}\label{timelikemkdvcurveflow}
\pos_{t}=-T_{xx} +\tfrac{3}{2}\eta(T_x,T_x)T
= -\pos_{xxx} +\tfrac{3}{2}\eta(\pos_{xx},\pos_{xx})\pos_x,
\quad 
\eta(\pos_x,\pos_x)=-1
\end{equation}
which is a timelike version of the non-stretching mKdV map equation
\cite{Anc06} in Minkowski space $\Mink{2}$.

\section{Spacelike curve flows and integrable systems in $\Mink{2}$}
\label{3dimspacelike}

In 3-dimensional Minkowski space $\Mink{2}$, 
let $\pos(x)=(X^0(x),X^1(x),X^2(x))$ be a spacelike curve, 
with the tangent vector $\pos_x$ satisfying 
\begin{equation}
\eta(\pos_x,\pos_x)>0
\label{spacelikeT}
\end{equation}
at every point $x$ on the curve. 
Hereafter we choose $x$ to be the proper-distance arclength parameter defined by
\begin{equation}
|\pos_{x}|=\sqrt{\eta(\pos_{x},\pos_{x})}=1 . 
\label{properdistance}
\end{equation} 
Then 
\begin{equation}
T=\pos_x, \quad\text{unit spacelike tangent vector}
\label{spaceliketangent}
\end{equation}
satisfies
\begin{equation}
\eta(T,T)=1 .
\label{spacelikeunitT}
\end{equation}
Hence the normal plane orthogonal to $T$ at each point on the curve 
is a Minkowski plane, $\Mink{1}$. 
The $x$-derivative of $T$ defines the principal normal vector $T_x$
which lies in this plane, 
due to $\eta(T,T_x)=0$. 
Thus $T_x$ can be either timelike, spacelike, or null. 
The case when $T_x$ is non-null is similar to the cases of 
timelike/spacelike curve flows in 2-dimensional Minkowski space 
discussed in \secref{2dimnonnullcase}. 
In contrast, the case when $T_x$ is null has no counterpart 
in 2-dimensional Minkowski space. 

We will consider the non-null case first and discuss the null case afterward.

\subsection{Spacelike curve flows with a non-null normal vector}
\label{spacelikenonnullnormal}
We begin by introducing the Lorentzian analog of a Frenet frame 
for spacelike curves whose principal normal vector $T_x$ 
is assumed to be non-null 
\begin{equation}
\eta(T_x,T_x)\neq 0
\label{nonnullprincipal}
\end{equation}
at every point on the curve. 
In this case, 
\begin{equation}
N=\kappa^{-1}T_x, \quad\text{unit timelike/spacelike normal vector}
\label{nonnullnormal}
\end{equation}
satisfies
\begin{equation}
\eta(T,N)=0,
\quad
\eta(N,N)=\mp 1
\label{nonnullunitN}
\end{equation}
where the function $\kappa(x)$ is given by 
\begin{equation}
\kappa=\mp\eta(T_x,N)= \sqrt{\mp\eta(T_x,T_x)}
\label{nonnullcurvinv}
\end{equation}
which is the Lorentzian curvature invariant of the curve. 
As shown by property \eqref{cross2}, 
the cross product of $T$ and $T_x$ yields a vector 
which is orthogonal to both $T$ and $T_x$ 
and has the norm 
$\eta(T\times T_x,T\times T_x)= -\eta(T_x,T_x)$
which is opposite in sign to the norm of $T_x$. 
Hence 
\begin{equation}
B=\kappa^{-1} T\times T_x, \quad\text{unit spacelike/timelike bi-normal vector}
\label{nonnullbinormal}
\end{equation}
satisfies
\begin{equation}
\eta(B,T)=\eta(B,N)=0,
\quad
\eta(B,B)=\pm 1 . 
\label{nonnullunitB}
\end{equation}
The triple of mutually orthogonal vectors 
\begin{equation}
\bmatr 
T\\  N \\ B
\ematr
\label{nonnullfrenetTNB}
\end{equation}
thereby defines a Frenet frame for a spacelike curve $\pos(x)$ 
with a non-null principal normal in $\Mink{2}$. 
This frame has the orientation 
\begin{equation}
\epsilon(T,N,B) = \pm 1
\end{equation}
in terms of the volume form. 

The Frenet equations of this frame \eqref{nonnullfrenetTNB} 
can be derived similarly to the case of timelike curves in \secref{3dimtimelikeframe},
with some changes of signs. 
First, from equation \eqref{nonnullnormal}, 
\begin{equation}
T_x=\kappa N
\label{nonnullTx}
\end{equation}
is the same. 
Next, 
from the $x$-derivative of equation \eqref{nonnullunitN}, 
we have 
\begin{equation}
N_x=\pm\kappa T+\tau B
\label{nonnullNx}
\end{equation}
for some function $\tau(x)$. 
An expression for $\tau$ can be obtained 
in terms of $T$ and its $x$-derivatives
by use of equations \eqref{nonnullnormal} and \eqref{nonnullbinormal}. 
This yields
\begin{equation}
\tau=\pm\eta(N_x,B)=\pm\kappa^{-2}\eta(T\times T_x,T_{xx})
\label{nonnulltorsinv}
\end{equation}
which is the Lorentzian torsion invariant of the curve. 
Last, from the $x$-derivative of equation \eqref{nonnullbinormal}
combined with equations \eqref{nonnullTx}, \eqref{nonnullNx}, 
and property \eqref{cross1}, 
we have
\begin{equation}
B_x=\tau N
\label{nonnullBx}
\end{equation} 
which has changed in sign. 

Therefore, the Frenet equations are given by
\begin{equation}
\bmatr 
T_x\\  N_x \\ B_x
\ematr=\bmatr 
0 & \kappa & 0\\ \pm\kappa & 0 & \tau \\ 0 & \tau
& 0
\ematr\bmatr 
T\\  N \\ B
\ematr
\label{nonnullfreneteqns}
\end{equation}
where $\bmatr 0 & \kappa & 0\\ \pm\kappa & 0 & \tau \\ 0 & \tau & 0 \ematr\in \mathfrak{so}(2,1)$ is the Cartan matrix,
which belongs to the Lie algebra $\mathfrak{so}(2,1)$ of the 
$SO(2,1)$ group of rotation and boost isometries in $\Mink{2}$.

A general frame for a spacelike curve in $\Mink{2}$ is related to 
this Frenet frame by the action of arbitrary $x$-dependent $SO(2,1)$ 
rotations and boosts applied to the vectors \eqref{nonnullfrenetTNB}. 
If the tangent vector $T$ is preserved as one of the frame vectors, 
then the resulting frame is given by 
applying a general $x$-dependent $SO(1,1)$ boost
on the normal vectors $N$ and $B$ in the $\Mink{1}$ normal plane of the curve. 
Geometrically, 
this transformation on the Frenet frame is a hyperbolic rotation,
yielding 
\begin{equation}
\E=\bmatr e_\parallel\\e_1 \\e_2 \ematr
=
\bmatr 
1&0           &0            \\
0&\cosh\theta &-\sinh\theta  \\
0&-\sinh\theta &\cosh\theta\\
\ematr
\bmatr T\\N\\B\\\ematr
=\bmatr 
T \\ (\cosh\theta) N-(\sinh\theta) B \\ -(\sinh\theta) N+(\cosh\theta)B
\ematr
\label{nonnulladaptedframe}
\end{equation}
where $\theta(x)$ is the hyperbolic rotation angle,
and where the frame vectors satisfy the orthogonality relations
\begin{gather}
\eta(e_\parallel,e_\parallel)=1,
\quad 
\eta(e_1,e_1)=-\eta(e_2,e_2)=\mp 1, 
\label{nonnull1}\\
\eta(e_\parallel,e_1)=\eta(e_\parallel,e_2)=\eta(e_1,e_2)=0
\label{nonnull2}
\end{gather}
and the cross product relations
\begin{equation}
e_1\times e_2=\pm e_\parallel,
\quad
e_\parallel\times e_1=e_2,
\quad
e_\parallel\times e_2=e_1 
\end{equation}
which follow from $\epsilon(e_0,e_1,e_2)=\epsilon(T,N,B) = \pm1$. 

Similarly to the case of timelike curves, 
this gauge freedom can be used to define a {\em Lorentzian parallel frame}
by the geometrical condition that the $x$-derivative of 
both normal vectors $e_1$ and $e_2$ is parallel to the tangent vector $e_\parallel$. 
Thus we require 
 \begin{equation}
e_1{}_x=u_1e_\parallel, 
\quad 
e_2{}_x=u_2e_\parallel
\label{nonnullparallel1}
\end{equation}
for some functions $u_1(x),u_2(x)$. 
To obtain $e_\parallel{}_x$, 
we substitute equation \eqref{nonnullparallel1}
into the $x$-derivative of equations \eqref{nonnull1} and \eqref{nonnull2}.
This yields the relations 
$\eta(e_\parallel{}_x,e_1)= -u_1$, 
$\eta(e_\parallel{}_x,e_2)= -u_2$, 
and $\eta(e_\parallel{}_x,e_\parallel)=0$,
which determine 
\begin{equation}
e_\parallel{}_x=\pm( u_1e_1 -u_2e_2) . 
\label{nonnullparallel2}
\end{equation}
Therefore, the Cartan matrix of a Lorentzian parallel frame 
for a spacelike curve with a non-null principal normal in $\Mink{2}$ 
is given by 
\begin{equation}
\E_x=\U\E
\label{nonnullEx}
\end{equation}
where 
\begin{equation}
\U=\bmatr 
0 & \pm u_1& \mp u_2\\ u_1 & 0 & 0\\ u_2 & 0 & 0
\ematr
\in \mk{so}(2,1) . 
\label{nonnullparallelC}
\end{equation}
This matrix belongs to the perp space of the stabilizer subalgebra 
$\mk{so}(1,1) \subset \mk{so}(2,1)$ of the frame vector $e_\parallel$. 
The perp space $\mk{so}(1,1)_\perp\simeq \Mink{1}$ is given by 
the span of the rotation subalgebra $\mk{so}(2)$ 
and another boost subalgebra $\mk{so}(1,1)$ 
which act in two orthogonal planes (one spacelike and one timelike) 
containing $e_\parallel$. 
In particular, 
$\mk{so}(1,1)\oplus \mk{so}(1,1)_\perp = \mk{so}(2,1)$
is a symmetric Lie algebra decomposition. 

To determine the hyperbolic rotation angle $\theta(x)$ 
under which a Frenet frame is transformed to a parallel frame, 
we first need the inverse transformation
\begin{equation}
\bmatr T\\N \\B\ematr
=
\bmatr 
e_\parallel \\
(\cosh\theta) e_1 +(\sinh\theta)e_2  \\
(\sinh\theta) e_1+(\cosh\theta) e_2 \\
\ematr . 
\label{nonnullinvtrans}
\end{equation}
Now, we take the $x$-derivative of the frame \eqref{nonnulladaptedframe} 
and substitute the Frenet equations \eqref{nonnullfreneteqns} 
followed by the inverse transformation \eqref{nonnullinvtrans}, 
which yields
\begin{equation}
\E_x=
\bmatr 
(\kappa \cosh\theta) e_1 +(\kappa\sinh\theta) e_2\\
\pm(\kappa \cosh\theta) e_\parallel -(\theta_x-\tau) e_2  \\
\mp(\kappa \sinh\theta) e_\parallel -(\theta_x-\tau) e_1 
\ematr . 
\end{equation}
Thus the condition \eqref{nonnullEx}--\eqref{nonnullparallelC}
can be achieved if (and only if)
\begin{equation}
\theta_x = \tau . 
\label{nonnullrotationcond}
\end{equation} 
The resulting frame given by equations \eqref{nonnulladaptedframe} 
and \eqref{nonnullrotationcond} thereby defines a parallel frame, 
where the components of its Cartan matrix are related to 
the curvature and torsion invariants $\kappa,\tau$ by 
\begin{equation}
u_1=\pm\kappa\cosh\theta=\pm\kappa \cosh\left(\txtint \tau dx\right),
\quad 
u_2=\mp\kappa\sinh\theta=\mp\kappa \sinh\left(\txtint \tau dx\right) . 
\end{equation}
These expressions are a Lorentzian counterpart of the well-known 
Hasimoto transformation \cite{Has} in Euclidean space. 

A parallel frame is unique up to $x$-independent (rigid) hyperbolic rotations 
\begin{equation}
\theta\rightarrow\theta+\phi
\end{equation}
where $\phi$ is constant.
Under these transformations, 
the tangent vector $e_\parallel$ is preserved,
while the normal vectors $e_1$ and $e_2$ are rigidly rotated
\begin{equation}
\begin{aligned}
& e_1\rightarrow (\cosh\phi)e_1 -(\sinh\phi)e_2,
\\
& e_2\rightarrow -(\sinh\phi)e_1+(\cosh\phi)e_2 . 
\end{aligned}
\label{nonnullrigidrotation}
\end{equation}
Any two parallel frames related by this hyperbolic rotation 
are gauge equivalent. 
The $SO(1,1)$ group of rigid hyperbolic rotations thereby defines 
the gauge (equivalence) group for parallel frames. 
Stated another way, 
a parallel frame for a spacelike curve with a non-null principal normal 
is determined only up to the action of this gauge group. 

\subsubsection{Inelastic Flow Equations}
We now consider spacelike inelastic curve flows $\pos(t,x)$ that locally preserve 
the proper distance normalization \eqref{propertime} of the arclength parameter
and the non-null signature \eqref{nonnullprincipal} of the principal normal.
Such flows are specified by a flow vector 
\begin{equation}\label{nonnullflow}
\pos_t=h_\parallel e_\parallel+h_1e_1+h_2e_2
\end{equation}
expressed in terms of a tangential component $h_\parallel$ 
and a pair of normal components $h_1,h_2$ 
with respect to the frame vectors $e_\parallel$, $e_1$, $e_2$. 

The parallel frame will be carried by the flow, 
such that the orthogonality relations \eqref{nonnull1}--\eqref{nonnull2}
are preserved. 
This implies that the $t$-derivative of the frame vectors 
$e_\parallel$, $e_1$, $e_2$ is given by 
\begin{gather}
e_\parallel{}_t=\pm(\omega_1e_1 -\omega_2e_2),
\\
e_1{}_t=\omega_1e_\parallel+\omega_0e_2,
\quad 
e_2{}_t=\omega_2e_\parallel +\omega_0e_1 .
\end{gather}
We can write these equations in the form
\begin{equation}
\E_t=\W \E
\label{nonnullEt}
\end{equation}
with the Cartan matrix
\begin{equation}
\W=\bmatr 
0 & \pm\omega_1& \mp\omega_2\\ \omega_1 & 0 & \omega_0\\ \omega_2 & \omega_0 & 0
\ematr
\in \mk{so}(2,1)
\label{nonnullflowC}
\end{equation}
which belongs to the Lie algebra of the $SO(2,1)$ group of 
rotation and boost isometries in $\Mink{2}$.

The flow equations \eqref{nonnullEt} 
and the Frenet equations \eqref{nonnullEx}
of the parallel frame are related by the compatibility condition 
$\partial_t(\E_x)=\partial_x(\E_t)$, 
which is equivalent to a zero curvature equation \eqref{timelikezerocurv}
relating the Cartan matrices $\W$ and $\U$. 
Substituting these matrices \eqref{nonnullparallelC} and \eqref{nonnullflowC} 
into equation \eqref{timelikezerocurv}, 
we obtain
\begin{gather}
u_{1t}=\omega_{1x}+u_2\omega_0,
\quad 
u_{2t}=\omega_{2x}+u_1\omega_0,
\label{nonnullzerocurv1}\\
\omega_{0x}=\mp(u_1\omega_2-\omega_1u_2) . 
\label{nonnullzerocurv2}
\end{gather}
There is a similar compatibility condition 
$\partial_x(\pos_t)=\partial_t(\pos_x)$
relating the flow vector \eqref{nonnullflow} 
and the tangent vector \eqref{spaceliketangent} 
of the curve,
which is given by equation \eqref{timelikezerotors}
using the notation \eqref{timelikeHe}. 
After substituting the matrices \eqref{nonnullparallelC} and \eqref{nonnullflowC} 
along with the vectors \eqref{timelikeHe}
into this equation, 
we find that its tangential and normal components yield 
\begin{gather}
h_{\parallel}{}_x=-h_1u_1-h_2u_2 , 
\label{nonnullzerotors1}\\
\omega_1=\pm h_1{}_x+u_1h_\parallel,
\quad 
\omega_2=\mp h_2{}_x+u_2h_\parallel . 
\label{nonnullzerotors2}
\end{gather} 

These compatibility equations 
\eqref{nonnullzerocurv1}, \eqref{nonnullzerocurv2}, 
\eqref{nonnullzerotors1}, \eqref{nonnullzerotors2} 
are the Cartan structure equations of the parallel frame,
describing all inelastic spacelike curve flows $\pos(t,x)$ 
with a non-null principal normal in $\Mink{2}$. 

\subsubsection{SO(1,1)-invariant formalism}
The gauge group for parallel frames consists of 
rigid $SO(1,1)$ hyperbolic rotations \eqref{nonnullrigidrotation}
acting on the normal vectors. 
Under the action of this group, 
the components of the Cartan matrix \eqref{nonnullparallelC} 
along the curve are transformed by 
\begin{equation}
\begin{aligned}
& u_1\rightarrow u_1\cosh\phi -u_2\sinh\phi=\pm\kappa \cosh(\theta+\phi), 
\\
& u_2\rightarrow -u_1\sinh\phi+u_2\cosh\phi=\mp\kappa \sinh(\theta+\phi) .
\end{aligned}
\label{nonnulluequiv}
\end{equation}
Similarly, the components of the Cartan matrix \eqref{nonnullflowC} 
along the flow are transformed by
\begin{gather}
\omega_0\rightarrow\omega_0,
\label{nonnullw0equiv}
\\
\begin{aligned}
& \omega_1\rightarrow\omega_1\cosh\phi -\omega_2\sinh\phi,
\\
& \omega_2\rightarrow -\omega_1\sinh\phi+\omega_2\cosh\phi.
\end{aligned}
\label{nonnullwequiv}
\end{gather}
Since the flow vector \eqref{nonnullflow} is gauge invariant, 
its normal and tangential components in a parallel frame 
are transformed by 
\begin{gather} 
h_\parallel\rightarrow h_\parallel,
\label{nonnullhparequiv}
\\
\begin{aligned}
& h_1\rightarrow h_1\cosh\phi+h_2\sinh\phi,
\\
& h_2\rightarrow h_1\sinh\phi+h_2\cosh\phi.
\end{aligned}
\label{nonnullhperpequiv}
\end{gather}
A comparison with the case of rigid $SO(2)$ rotations 
in \secref{SO(2)invariance} 
suggests that we introduce a hyperbolic analog of complex variables. 

Let $j$ denote a unit hyperbolic number (also called a split-complex number)
defined by 
\begin{equation}
j^2=+1,
\quad 
\bar{j}=-j 
\end{equation}
in analogy to the complex imaginary number $i^2=-1$, $\bar{i}=-i$. 
The hyperbolic version of the Euler relation is 
\begin{equation}
\exp(j\theta)=\cosh\theta+j\sinh\theta . 
\end{equation}
If we write
\begin{gather}
u=u_1+ju_2=\pm\kappa \exp(-j\theta)=\pm\kappa \exp(-j\txtint\tau dx),
\quad
\omega=\omega_1+j\omega_2,  
\\
h_\perp=\pm(h_1-jh_2) , 
\end{gather}
then the rigid $SO(1,1)$ gauge transformations \eqref{nonnulluequiv}--\eqref{nonnullhperpequiv}
become a hyperbolic phase rotation
\begin{equation}
u\rightarrow u\exp(-j\phi),
\quad
\omega\rightarrow \omega\exp(-j\phi),
\quad 
h_\perp\rightarrow h_\perp\exp(-j\phi).
\end{equation}

We can now write the evolution equations in a $SO(1,1)$-invariant form:
\begin{gather}
u_t=\omega_x+ju\omega_0,
\label{nonnullu}\\
\omega_{0x}=\mp\Im(\bar{u}\omega),
\label{nonnullw0}\\
h_\parallel{}_x=\mp\Re( \bar{u} h_\perp),
\label{nonnullh}\\
\omega= h_\perp{}_x+uh_\parallel.
\label{nonnullw}
\end{gather}
Since $u$ is determined by the curve 
only up to rigid hyperbolic phase rotations \eqref{nonnulluequiv}, 
this split-complex variable $u$ is a Hasimoto variable 
which geometrically represents a {\em $SO(1,1)$-covariant} of the curve,
in contrast to the invariants $\kappa,\tau$
which are determined uniquely by the curve. 

This system \eqref{nonnullu}--\eqref{nonnullw} has an operator formulation,
encoding a triple of $SO(1,1)$-invariant Hamiltonian operators, as follows.

\begin{thm}
\label{nonnullHJthm}
For inelastic flows of spacelike curves with a non-null principal normal 
in $\Mink{2}$, 
the curve covariant $u$ satisfies the $SO(1,1)$-invariant system 
\begin{align}
& u_t=D_x\omega \mp ju D_x^{-1}\Im(\omega \bar u) = \Hop(\omega)
\label{nonnullueqn}
\\
& \omega= D_{x}h_\perp \mp uD_{x}^{-1}\Re(\bar u h_\perp)=\Jop(h_\perp)
\label{nonnullweqn}
\end{align}
where 
\begin{align}
& \Hop=D_x \pm juD_{x}^{-1}\Im(u\Cop)
\label{nonnullH}\\
& \Jop=D_{x} \mp uD_{x}^{-1}\Re(u\Cop)
\label{nonnullJ}
\end{align}
are a pair of compatible, $SO(1,1)$-invariant
Hamiltonian and symplectic operators,
and $\Cop$ denotes the complex conjugation operator. 
Moreover, the operators $\Hop$ and $\Jop$ are related by 
\begin{equation}
\Hop\Iop^{-1}= \Iop\Jop,
\quad
\Jop\Iop= \Iop^{-1}\Hop,
\quad
\Iop =j 
\label{nonnullI}
\end{equation}
where $\Iop$ is is a Hamiltonian operator 
compatible with $\Hop$ and $\Jop$. 
Composition of these operators yields the $SO(1,1)$-invariant evolution equation
\begin{equation}\label{nonnulluflow}
u_t=\Rop^2(h_\perp)
\end{equation}
for $u$ in terms of the normal component $h_\perp$ of the flow, 
where 
\begin{equation}
\Rop=\Hop\Iop^{-1} =\Iop\Jop = j(D_x \mp uD_x^{-1}\Re(u\Cop))
\end{equation}
is a hereditary recursion operator. 
\end{thm}

This theorem is analogous to the similar result 
for timelike inelastic curve flows in \secref{SO(2)invariance}
and has a similar proof by replacing $i$ with $j$. 

An operator $\Dop$ here is Hamiltonian iff 
it defines an associated Poisson bracket \eqref{poisson} 
having the same structure as in the case of timelike inelastic curve flows, 
with $j$ in place of $i$. 
Similarly, based on Magri's general theorem \cite{Mag}, 
the $SO(1,1)$ invariance of the pair of compatible Hamiltonian operators 
$\Hop$ and $\Iop$
can be used here to derive a hierarchy of flows starting from a root flow 
$u_t=ju$ corresponding to the generator of $SO(1,1)$ hyperbolic phase-rotations on $u$. 
This leads to the following main result. 

\begin{thm}
\label{nonnullhierarchy}
There is a hierarchy of integrable tri-Hamiltonian flows on $u(t,x)$
given by
\begin{equation}
\begin{aligned}
u_t=\Rop^{n+1}(ju) & 
= \Iop(\delta \mk{H}^{(n+1)}/\delta\bar u) 
=\Hop(\delta \mk{H}^{(n)}/\delta\bar u) , 
\quad 
n=0,1,2,\ldots 
\\& 
=\Eop(\delta \mathfrak{H}^{(n-1)}/\delta\bar u), 
\quad 
n=1,2,\ldots 
\end{aligned}
\label{nonnulltrihamil}
\end{equation}
(called the {\em $+n$ flow})
in terms of Hamiltonians $\mk{H}^{(n)}=\int H^{(n)} dx$
where 
\begin{equation}
\begin{aligned}
& \Iop =j,
\quad
\Hop=D_x \pm juD_x^{-1}\Im(u\Cop),
\\
& \Eop = \Rop\Hop= j(D_x{}^2 \mp|u|^2 \pm uD_x^{-1}\Re(u_x\Cop)\pm ju_xD_x^{-1}\Im(u\Cop))
\end{aligned}
\end{equation}
are compatible Hamiltonian operators, 
and where 
\begin{equation}
H^{(n)}=
2(1+n)^{-1} D^{-1}_x\Im(\bar u(j\Hop)^{n+1} u), 
\quad 
n=0,1,2,\ldots
\label{nonnullHam}
\end{equation}
are local Hamiltonian densities. 
\end{thm}

The expression for the Hamiltonians \eqref{nonnullHam} in this theorem 
arises from a scaling formula derived in \Ref{Anc03}. 

The $+1$ flow in the hierarchy \eqref{nonnulltrihamil} is explicitly given by 
\begin{equation}\label{nonnullnls}
u_t=j(u_{xx}\mp\tfrac{1}{2}|u|^{2}u)
\end{equation}
which is a variant of the defocusing NLS equation
with $j$ in place of $i$. 
This equation \eqref{nonnullnls} has the explicit tri-Hamiltonian structure
\begin{equation}\label{nonnullnlsnlshamil}
u_t = \Iop(\delta\mk{H}^{(2)}/\delta\bar u) 
=\Hop(\delta \mk{H}^{(1)}/\delta\bar u) 
=\Eop(\delta\mk{H}^{(0)}/\delta\bar u)  
\end{equation}
in terms of the Hamiltonian densities 
\begin{equation}\label{nonnullH012}
H^{(0)} = |u|^2,
\quad
H^{(1)} = \Im(u_x\bar u),
\quad
H^{(2)} = - |u_x|^2 \mp \tfrac{3}{4} |u|^4
\end{equation}
(modulo trivial densities given by total $x$-derivatives). 
Similarly, the $+2$ flow in the hierarchy \eqref{nonnulltrihamil} is given by 
\begin{equation}\label{nonnullmkdv}
u_t=u_{xxx}\mp\tfrac{3}{2}|u|^{2}u_x
\end{equation}
which is a variant of the complex defocusing mKdV equation,
where $j$ replaces $i$. 
It has the tri-Hamiltonian structure
\begin{equation}\label{nonnullmkdvhamil}
u_t = \Iop(\delta\mk{H}^{(3)}/\delta\bar u) 
=\Hop(\delta \mk{H}^{(2)}/\delta\bar u) 
=\Eop(\delta\mk{H}^{(1)}/\delta\bar u)  
\end{equation}
where $\mk{H}^{(1)}$ and $\mk{H}^{(2)}$ are given by 
the Hamiltonian densities \eqref{nonnullH012}, 
while $\mk{H}^{(3)}$ is given by the Hamiltonian density 
\begin{equation}\label{nonnullH3}
H^{(3)} = \Im(u_x\bar u_{xx}) \pm\tfrac{3}{4}|u|^2\Im(u\bar u_x)
\end{equation}
(modulo trivial densities given by total $x$-derivatives). 

In exactly the same way as for timelike inelastic curve flows, 
here the entire hierarchy \eqref{nonnulltrihamil} of flows 
corresponds to a hierarchy of commuting vector fields 
\begin{equation}\label{spacelikesymms}
\X^{(n)} = \Rop^n(ju)\partial_u,
\quad
n=0,1,2,\ldots
\end{equation}
with the root vector field being the generator of hyperbolic phase-rotations
$\X^{(0)} = ju\partial_u$. 
All of the vector fields in the hierarchy are bi-Hamiltonian, 
with 
$\pr\X^{(n)}\mk{G} = \{\mk{G},\mk{H}^{(n)}\}_{\Hop}= \{\mk{G},\mk{H}^{(n+1)}\}_{\Iop}$
for $n\geq0$, 
where $\mk{H}^{(n)}=\int H^{(n)}dx$ is the functional 
with the Hamiltonian density \eqref{nonnullHam},
while each vector field for $n\geq 1$ is also tri-Hamiltonian, 
$\pr\X^{(n)}\mk{G} = \{\mk{G},\mk{H}^{(n-1)}\}_{\Eop}$. 
Since the entire hierarchy is commuting, 
every Hamiltonian vector field \eqref{spacelikesymms} 
is the generator of a symmetry 
for the defocusing NLS equation \eqref{nonnullnls}
as well as the defocusing mKdV equation \eqref{nonnullmkdv}, 
and every associated Hamiltonian \eqref{nonnullHam} 
is a conserved density 
for both of these equations. 

\begin{prop}
Each tri-Hamiltonian flow in the hierarchy \eqref{nonnulltrihamil} 
can be written in the form \eqref{nonnulluflow} as given by 
\begin{equation}
h_\perp=\Rop^{n-1}(ju), 
\quad 
n=1,2,\ldots . 
\end{equation}
Consequently, 
the $+n$ flow for $n\geq1$ determines an inelastic timelike curve flow 
\begin{equation}\label{nonnullcurveflow}
\pos_t=h_\parallel^{(n)}e_\parallel + h_1^{(n)}e_1 + h_2^{(n)}e_2
\quad
n=1,2,\ldots
\end{equation}
whose tangential and normal components are given by 
\begin{equation}\label{nonnullflowh}
h_\parallel^{(n)}= \tfrac{1}{2}(n-1) H^{(n-2)},
\quad
h_1^{(n)}-j h_2^{(n)} =\pm \Rop^{n-1}(ju), 
\quad
n=1,2,\ldots . 
\end{equation}
\end{prop}

These equations of motions \eqref{nonnullcurveflow} 
can be written in an explicit form by use of the following variant of 
a complex parallel-frame notation:
\begin{equation}
e_\parallel =T, 
\quad 
e_\perp=e_1+je_2 = e^{-j\theta}(N+jB)
\label{nonnullcomplexE}
\end{equation}
satisfying
\begin{equation}
e_\parallel\times e_\perp= je_\perp, 
\quad
e_\perp\times \bar e_\perp= \mp j2e_\parallel
\label{nonnullcrossE}
\end{equation}
and 
\begin{equation}
\eta(e_\parallel, e_\parallel) =1, 
\quad
\eta(e_\perp,\bar e_\perp)= \mp 2, 
\quad
\eta(e_\parallel,e_\perp)=\eta(e_\perp,e_\perp)=0 . 
\label{nonnullcomplexEnorms}
\end{equation}
The action of the gauge group \eqref{nonnullrigidrotation}
on these frame vectors consists of rigid hyperbolic phase rotations
\begin{equation}\label{nonnullcomplexequiv}
e_\parallel \rightarrow e_\parallel,
\quad 
e_\perp \rightarrow e^{-j\phi}e_\perp
\end{equation}
where $\phi$ is constant. 

The Frenet equations \eqref{nonnullEx} of the underlying parallel frame 
become
\begin{equation}
e_\parallel{}_x=\pm\Re(\bar u e_\perp) ,\quad
e_\perp{}_x=ue_\parallel , 
\label{nonnullcomplexEx}
\end{equation}
while the flow equations \eqref{nonnullEt} are similarly given by 
\begin{equation}
e_\parallel{}_t=\pm\Re(\bar\omega e_\perp) ,
\quad
e_\perp{}_t=\omega e_\parallel +j\omega_0 e_\perp . 
\label{nonnullcomplexEt}
\end{equation}

The equation of motions \eqref{nonnullcurveflow} for the curve 
then take the form 
\begin{equation}\label{nonnullcomplexcurveflow}
\pos_t=h_\parallel^{(n)}e_\parallel \pm\Re(\bar h_\perp^{(n)} e_\perp), 
\quad
n=1,2,\ldots
\end{equation}
where the tangential component $h_\parallel^{(n)}$
and the normal component $h_\perp^{(n)}=\pm(h_1^{(n)}-jh_2^{(n)})$
are functions of the curve covariant $u$, 
the conjugate covariant $\bar u$,
and their $x$-derivatives, 
as given by the expressions \eqref{nonnullflowh}. 
Since $u\rightarrow e^{-j\phi}u$ and $\bar u\rightarrow e^{j\phi}\bar u$
under the gauge group \eqref{nonnullcomplexequiv}, 
the tangential component is gauge invariant 
while the normal component is gauge equivariant, 
\begin{equation}
h_\parallel^{(n)} \rightarrow h_\parallel^{(n)},
\quad
h_\perp^{(n)} \rightarrow e^{-j\phi} h_\parallel^{(n)} . 
\end{equation}
Consequently, 
each equation of motion \eqref{nonnullflow} is invariant under 
the isometry group $ISO(2,1)$ of $\Mink{2}$
and thus describes a geometric non-stretching motion of the curve.

The curve flow $n=1$ 
corresponding to the defocusing NLS equation \eqref{nonnullnls}
is determined by the components
\begin{equation}
h_\parallel^{(1)}=0,
\quad
h_\perp^{(1)}=ju . 
\end{equation}
Substituting these expressions into the $n=1$ equation of motion \eqref{nonnullcomplexcurveflow}, 
we have 
$\pos_t = \pm\Re(-j\bar u e_\perp)
=\mp\Re(\bar u e_\parallel\times e_\perp)
= -e_\parallel\times e_\parallel{}_x$
by using the cross product identity \eqref{nonnullcrossE} 
and the Frenet equations \eqref{nonnullcomplexEx}. 
Thus we obtain 
\begin{equation}\label{nonnullnlscurveflow}
-\pos_{t}=T\times T_x = \pos_x\times\pos_{xx},
\quad 
\eta(\pos_x,\pos_x)=-1,
\quad
\eta(\pos_{xx},\pos_{xx})\neq 0
\end{equation}
which is a spacelike version of the vortex filament equation in 
Minkowski space $\Mink{2}$. 

The curve flow $n=2$ 
corresponding to the defocusing mKdV equation \eqref{nonnullmkdv}
has the components
\begin{equation}
h_\parallel^{(2)}=\mp\tfrac{1}{2}|u|^2,
\quad
h_\perp^{(2)}=u_x . 
\end{equation}
Substituting these expressions into the $n=2$ equation of motion \eqref{nonnullflow}, 
we find 
$\pos_t = \pm\Re(\bar u_x e_\perp) \mp\tfrac{1}{2}|u|^2 e_\parallel$.
We simplify the term 
$\Re(\bar u_x e_\perp)= \Re(\bar u e_\perp)_x - \Re(\bar u e_\perp{}_x)
= \pm e_\parallel{}_{xx} - |u|^2 e_\parallel$
by using the Frenet equations \eqref{nonnullcomplexEx}. 
Next we note 
$\eta(e_\parallel{}_x,e_\parallel{}_x) 
=\eta(\pm\Re(\bar u e_\perp),\pm\Re(\bar u e_\perp))
= \tfrac{1}{2} \eta(\bar u e_\perp,u\bar e_\perp)= \mp|u|^2$
which follows from the relations \eqref{nonnullcomplexEnorms}. 
This yields 
$\pos_t = e_\parallel{}_{xx} \mp\tfrac{3}{2}|u|^2 e_\parallel$,
and thus we obtain 
\begin{equation}\label{nonnullmkdvcurveflow}
\pos_{t}=T_{xx} \mp\tfrac{3}{2}\eta(T_x,T_x)T
= \pos_{xxx} \mp\tfrac{3}{2}\eta(\pos_{xx},\pos_{xx})\pos_x,
\quad 
\eta(\pos_x,\pos_x)=1, 
\quad
\eta(\pos_{xx},\pos_{xx})\neq 0
\end{equation}
which is a spacelike version of the non-stretching mKdV map equation
\cite{Anc06} in Minkowski space $\Mink{2}$.

\subsection{Spacelike curve flows with a null normal vector}
\label{spacelikenullnormal}
We now consider spacelike curves whose principal normal vector $T_x$ 
is assumed to be null 
\begin{equation}
\eta(T_x,T_x) =0
\label{nullprincipal}
\end{equation}
at every point on the curve. 

In contrast to the case when the principal normal is non-null, 
here the vectors $T_x$ and $T\times T_x$ are parallel, 
by the following argument. 
We note 
$\epsilon(T\times T_x,T,T_x)=\eta(T\times T_x,T\times T_x)
=-\eta(T,T)\eta(T_x,T_x)=0$ 
from the properties \eqref{3dimdual}, \eqref{3dimcross}, and \eqref{cross2}. 
Antisymmetry of $\epsilon$ thereby implies $T\times T_x=a T + b T_x$
for some functions $a(x)$, $b(x)$. 
Then, since $T\times T_x$ is orthogonal to $T$ from property \eqref{cross2}, 
we find 
$0=\eta(T,T\times T_x)=a$ due to $\eta(T,T_x)=0$, 
and hence we have $T\times T_x= b T_x$. 

To define a frame, 
we therefore need another vector, linearly independent of $T$ and $T_x$. 
Similarly to the case of null curves in 2 dimensions, 
it is natural to use a null vector 
on the opposite side of the lightcone in the normal plane $\Mink{1}$
(which is orthogonal to $T$). 
If $\vec v$ is a null vector in the plane $\Mink{1}$, 
let $\Nop$ be a linear map that produces a null vector $\Nop(\vec v)$
such that $\eta(\Nop(\vec v),\vec v) = -1$. 
Recall, from \secref{2dimnullcase}, 
that the null vectors $\vec v$ and $\Nop(\vec v)$ 
are spatial reflections of each other with respect to the timelike vector 
$\vec v+\Nop(\vec v)$ in $\Mink{1}$,
and that any change in the normalization $\eta(\Nop(\vec v),\vec v) = -1$ 
only produces a scaling of the null vector $\Nop(\vec v)$. 

A frame for a spacelike curve $\pos(x)$ 
with a null principal normal in $\Mink{2}$ 
can then be defined by 
\begin{subequations}\label{nullframe}
\begin{align}
& e_\parallel=\pos_x=T, 
\quad
\text{spacelike tangent vector}
\label{nullT}
\\
& e_{+}=\pos_{xx}=T_x, 
\quad
\text{null normal vector}
\label{nullTx}
\\
& e_{-} = \Nop(\pos_{xx}) = \Nop(T_x), 
\quad
\text{null opposite vector}
\label{nullopp}
\end{align}
\end{subequations}
with 
\begin{align}
& \eta(e_{\pm},e_\parallel)=0
\label{null1}
\\
& \eta(e_{\pm},e_{\pm})=0
\label{null2}
\\
& \eta(e_{+},e_{-})=-1
\label{null3}
\end{align}
where the properties \eqref{null1}--\eqref{null3} uniquely determine 
$e_{-}$ when $e_{+}$ and $e_\parallel$ are given.
Let 
\begin{equation}
\varepsilon =-\tfrac{1}{2}\epsilon(e_{+},e_{-},e_\parallel)
= \epsilon(e_0,e_1,e_\parallel), 
\quad
|\varepsilon|=1
\end{equation}
denote the orientation (equal to $+1$ or $-1$) of the frame vectors
as given in terms of the volume form,
where $e_0= \tfrac{1}{\sqrt{2}}(e_{+} + e_{-})$ is a unit timelike vector
and $e_1= \tfrac{1}{\sqrt{2}}(e_{+} - e_{-})$ is a unit spacelike vector
determined by the null normal vectors in the frame. 

The Frenet equations for this frame \eqref{nullframe} are 
easily derived by the same steps used 
in the case of null curves in 2 dimensions. 
First, from \eqref{nullTx} we have 
\begin{equation}
e_\parallel{}_x= e_+ . 
\label{nulleparx}
\end{equation}
Next, from the $x$-derivative of equations \eqref{null1} and \eqref{null2},
we have
$0=\eta(e_{+}{}_x,e_{+})$ and $0=\eta(e_{+}{}_x,e_\parallel)$ 
after using equation \eqref{nulleparx}. 
Thus we get 
\begin{equation}
e_{+x}=\sigma e_+
\label{nulleplusx}
\end{equation}
for some function $\sigma(x)$. 
Then, similarly we obtain 
$\eta(e_{-}{}_x,e_{-})=0$, $\eta(e_{-}{}_x,e_\parallel)=1$, 
$\eta(e_{-}{}_x,e_{+})=\sigma$, 
by using equations \eqref{null3}, \eqref{nulleparx}, and \eqref{nulleplusx}. 
This yields
\begin{equation}
e_{-x}=e_\parallel-\sigma e_- . 
\label{nulleminusx}
\end{equation}
Therefore, the Frenet equations are given by 
\begin{equation}
\bmatr 
e_{\parallel }\\  e_{+} \\ e_{-}
\ematr_{x}=\bmatr 
0 & 1 & 0\\
0 & \sigma &0 \\
1 & 0 & -\sigma
\ematr\bmatr 
e_\parallel\\  e_+ \\ e_-
\ematr
\label{nullfreneteqs}
\end{equation}
where the Cartan matrix $\bmatr  0 & 1 & 0\\ 0 & \sigma &0 \\ 1 & 0 & -\sigma \ematr\in \mk{so}(2,1)$ of the frame \eqref{nullframe}
belongs to the Lie algebra of the $SO(2,1)$ group of 
rotation and boost isometries in $\Mink{2}$.
These equations \eqref{nullfreneteqs} are preserved 
if the normalization \eqref{null3} of the null frame is changed. 
Therefore, $\sigma$ geometrically represents 
a Lorentzian curvature invariant of the curve. 

A general frame for a spacelike curve in $\Mink{2}$ is related to 
this Frenet frame by the action of arbitrary $x$-dependent $SO(2,1)$ 
rotations and boosts applied to the vectors \eqref{nullframe}. 
If the tangent vector $T$ is preserved as one of the frame vectors, 
then the resulting frame is given by 
applying a general $x$-dependent $SO(1,1)$ boost
on the null normal vectors $e_{\pm}$ in the $\Mink{1}$ normal plane of the curve.
This yields the boosted frame 
\begin{equation}
\E=\bmatr 
e_\parallel\\ \tilde{e}_+ \\\tilde{e}_-
\ematr
=\bmatr 
e_\parallel \\ \exp(\theta)e_+ \\ \exp(-\theta)e_-
\ematr
\label{nulladaptedframe}
\end{equation}
where the boost acts as a scaling, 
parameterize by $\theta(x)$, 
and where the frame vectors satisfy the orthogonality relations
\begin{gather}
\eta(e_\parallel,\tilde{e}_{\pm})=\eta(\tilde{e}_{\pm},\tilde{e}_{\pm})=0
\label{}\\
\eta({e}_\parallel,{e}_\parallel)=-\eta(\tilde{e}_+,\tilde{e}_-)=1
\label{}
\end{gather}
and the cross product relations
\begin{equation}
\tilde{e}_+\times\tilde{e}_-=-2\varepsilon e_\parallel, 
\quad
e_\parallel\times\tilde{e}_{\pm}=\pm2\varepsilon e_{\pm} . 
\end{equation}
To derive the Cartan matrix of the boosted frame \eqref{nulladaptedframe},
we first take the $x$-derivative of the frame vectors, 
and then we substitute the Frenet equations \eqref{nullfreneteqs}
followed by 
\begin{equation}
\bmatr 
e_\parallel\\ e_+ \\e_-
\ematr
=
\bmatr 
{e}_\parallel     \\
\exp(-\theta) \tilde{e}_+  \\
\exp(\theta) \tilde{e}_-\\
\ematr
\end{equation}
which yields 
\begin{equation}
\bmatr 
e_\parallel\\ \tilde{e}_+ \\\tilde{e}_-
\ematr_x
= 
\bmatr 
0& \exp( -\theta) &0 \\
0&\sigma+\theta_x &0  \\
\exp(-\theta)& 0 & -\theta_x-\sigma  
\ematr
\bmatr 
{e}_\parallel\\ \tilde{e}_+ \\ \tilde{e}_-
\ematr . 
\label{nulladaptedframeEx}
\end{equation}
We now see that, similarly to the case of a spacelike curve
with a non-null principal normal, 
here the gauge freedom in equation \eqref{nulladaptedframeEx}
can be used to define a Lorentzian parallel frame
by the geometrical condition that the $x$-derivative of 
each null normal vector $e_{\pm}$ is parallel to the tangent vector $e_\parallel$.
This condition can be achieved if (and only if)
\begin{equation}
\theta_x = -\sigma . 
\label{nullboostcond}
\end{equation} 
The resulting frame given by equations \eqref{nulladaptedframe} 
and \eqref{nullboostcond} is a parallel frame
whose Cartan matrix is given by 
\begin{equation}
\E_x=\U\E
\label{nullEx}
\end{equation}
where 
\begin{equation}
\U=\bmatr 
0 & u & 0 \\ 0 & 0 & 0\\ u & 0 & 0
\ematr
\in \mk{so}(2,1)
\label{nullparallelC}
\end{equation}
with 
\begin{equation}
u= \exp(-\theta)=\exp\left(-\txtint \sigma dx\right) . 
\end{equation}
Note that the Cartan matrix \eqref{nullparallelC} has one fewer component 
than in the case of a spacelike curve with a non-null principal normal,
since here the principal normal is constrained to lie on the lightcone. 
In particular, this Cartan matrix belongs to a 1-dimensional subalgebra
consisting of a boost combined with a rotation, 
which lies in a subspace of the perp space $\mk{so}(1,1)_\perp\simeq \Mink{1}$
of the stabilizer subalgebra $\mk{so}(1,1) \subset \mk{so}(2,1)$ 
of the frame vector $e_\parallel$. 

A parallel frame is unique up to $x$-independent (rigid) boosts given by 
\begin{equation}
\theta\rightarrow\theta+\phi
\end{equation}
where $\phi$ is constant.
Under these transformations, 
the tangent vector $e_\parallel$ is preserved,
while the null normal vectors $\tilde{e}_{\pm}$ are rigidly scaled 
\begin{equation}
\tilde{e}_{+}\rightarrow \exp(\phi)\tilde{e}_{+}, 
\quad
\tilde{e}_{-}\rightarrow \exp(-\phi)\tilde{e}_{-} . 
\label{nullrigidrotation}
\end{equation}
Any two parallel frames related by this transformation are gauge equivalent,
and thus the $SO(1,1)$ group of these rigid boosts thereby defines 
the gauge (equivalence) group for parallel frames. 
Stated another way, 
a parallel frame for a spacelike curve with a null principal normal 
is determined only up to the action of this gauge group. 

\subsubsection{Inelastic Flow Equations}
Let $\pos(t,x)$ be a spacelike inelastic curve flow that locally preserves
the proper distance normalization \eqref{propertime} of the arclength parameter
and the null signature \eqref{nullprincipal} of the principal normal.
Such flows are specified by a flow vector 
\begin{equation}\label{nullflow}
\pos_t=h_\parallel e_\parallel+h_{+}\tilde{e}_{+}+h_{-}\tilde{e}_{-}
\end{equation}
expressed in terms of a tangential component $h_\parallel$ 
and a pair of normal components $h_{\pm}$ 
with respect to the frame vectors $e_\parallel$, $\tilde{e}_{\pm}$. 

The parallel frame will be carried by the flow, 
such that the orthogonality relations \eqref{null1}--\eqref{null3}
are preserved. 
This implies that the $t$-derivative of the frame vectors 
$e_\parallel$, $\tilde{e}_{\pm}$ is given by 
\begin{gather}
e_\parallel{}_t=\omega_+ \tilde{e}_++\omega_- \tilde{e}_- , 
\\
\tilde{e}_{+}{}_t=\omega_-{e}_\parallel+\omega_0\tilde{e}_+,
\quad
\tilde{e}_{-}{}_t=\omega_+{e}_\parallel-\omega_0\tilde{e}_- . 
\end{gather}
We can write these equations in the form
\begin{equation}
\E_t=\W \E
\label{nullEt}
\end{equation}
with the Cartan matrix
\begin{equation}
\W=\bmatr 
0 & \omega_+& \omega_-\\ \omega_- &\omega_0& 0 \\  \omega_+ & 0 &-\omega_0
\ematr
\in \mk{so}(2,1)
\label{nullflowC}
\end{equation}
which belongs to the Lie algebra of the $SO(2,1)$ group of 
rotation and boost isometries in $\Mink{2}$.

Under the action of the gauge group of the parallel frame, 
the Cartan matrix \eqref{nullparallelC} along the curve is scaled by 
\begin{equation} 
u\rightarrow \exp(-\phi)u
\end{equation}
while the Cartan matrix \eqref{nullflowC} along the flow 
is similarly scaled by 
\begin{gather}
\omega_0\rightarrow\omega_0,
\\
\omega_\pm\rightarrow \exp(\mp\phi)\omega_\pm . 
\end{gather}
Since the flow vector \eqref{nullflow} is gauge invariant, 
its normal and tangential components also scale by 
\begin{gather}
h_\parallel\rightarrow h_\parallel,
\\
h_\pm\rightarrow  \exp(\mp\phi)h_\pm . 
\end{gather}

The flow equations \eqref{nullEt} 
and the Frenet equations \eqref{nullEx}
of the parallel frame are related by the compatibility condition 
$\partial_t(\E_x)=\partial_x(\E_t)$, 
which is equivalent to a zero curvature equation \eqref{timelikezerocurv}
relating the Cartan matrices $\W$ and $\U$. 
Substituting these matrices \eqref{nullparallelC} and \eqref{nullflowC} 
into equation \eqref{timelikezerocurv}, 
we obtain
\begin{gather}
\omega_-{}_x=0 , 
\label{nullzerocurv1}\\
\omega_0{}_x=-u \omega_- , 
\label{nullzerocurv2}\\
u_t=\omega_+{}_x-u\omega_0 . 
\label{nullzerocurv3}
\end{gather}
A similar compatibility condition 
$\partial_x(\pos_t)=\partial_t(\pos_x)$
relates the flow vector \eqref{nullflow} 
and the tangent vector \eqref{spaceliketangent} 
of the curve,
as formulated by equation \eqref{timelikezerotors}
using the notation \eqref{timelikeHe}. 
After substituting the matrices \eqref{nullparallelC} and \eqref{nullflowC} 
along with the vectors \eqref{timelikeHe}
into this equation, 
we find that its tangential and normal components yield 
\begin{gather}
h_{\parallel}{}_x=-uh_- , 
\label{nullzerotors1}\\
\omega_-= h_-{}_x , 
\label{nullzerotors2}\\
\omega_+= h_+{}_x+uh_\parallel .
\label{nullzerotors3}
\end{gather} 

These compatibility equations 
\eqref{nullzerocurv1}--\eqref{nullzerocurv3}
and \eqref{nullzerotors1}--\eqref{nullzerotors3} 
are the Cartan structure equations of the parallel frame,
describing all inelastic spacelike curve flows $\pos(t,x)$ 
with a null principal normal in $\Mink{2}$. 
Compared to all of the previous cases, 
here the system \eqref{nullzerocurv1}--\eqref{nullzerotors3} 
has a quite different operator structure, as follows.

From equations \eqref{nullzerocurv1} and \eqref{nullzerotors2}, 
we have 
\begin{equation}
h_- = a x + b, 
\quad
\omega_- = a
\end{equation}
for some functions $a(t)$, $b(t)$. 
Next, equation \eqref{nullzerocurv2} yields
\begin{equation}
\omega_0=-av,
\quad
v=\txtint u\; dx
\end{equation}
and then equation \eqref{nullzerotors1} similarly yields
\begin{equation}
h_\parallel=-(ax+b)v +a w,
\quad
w=\txtint v\; dx . 
\end{equation}
Substituting these expressions into equation \eqref{nullzerotors3}, 
we get
\begin{equation}
\omega_+= h_+{}_x+(a w-(ax+b)w_x)w_{xx} . 
\end{equation}
Finally, from equation \eqref{nullzerocurv3}, 
we obtain
\begin{equation}
u_t= w_{txx} = (h_++aww_x-\tfrac{1}{2}(ax+b)w_x^2)_{xx}
\end{equation}
which directly implies
\begin{equation}
w_t = h_++aww_x-\tfrac{1}{2}(ax+b)w_x^2 +cx+d
\end{equation}
for some functions $c(t)$, $d(t)$. 
Hence we have derived the following result. 

\begin{thm}
\label{nullthm}
For inelastic flows of spacelike curves with a null principal normal 
in $\Mink{2}$, 
the curve invariant $u$ satisfies the system 
\begin{equation}\label{nullut}
w_t = h_++aww_x-\tfrac{1}{2}(ax+b)w_x^2 +cx+d,
\quad
w_x=v, 
\quad
v_x=u
\end{equation}
for $w$ in terms of the normal component $h_+$ of the flow,
where $a(t)$, $b(t)$, $c(t)$, $d(t)$ are arbitrary functions. 
\end{thm}

We will now restrict attention to the class of flows that 
exhibit invariance under $x$-translations. This implies $a=c=0$, 
so thus the system \eqref{nullut} reduces to 
\begin{align}
& w_t +\tfrac{1}{2}w_x{}^2 = h_+ , 
\label{nullwt}
\\
& w_{xx}=u , 
\label{nullwpotential}
\end{align}
or equivalently 
\begin{align}
& v_t +v v_x = D_xh_+ , 
\label{nullvt}
\\
& v_x=u , 
\label{nullvpotential}
\end{align}
after a transformation 
$t\rightarrow \tilde t= \txtint b dt$
and $h_+\rightarrow \tilde{h}_+=(h_+ +d)/b$, 
when $b\neq 0$. 
The form of this system \eqref{nullvt}--\eqref{nullvpotential}
is closely related to both Burgers' equation and the KdV equation. 
Consider, first, the flow generated by 
\begin{equation}\label{nullburgersflow}
h_+=-v_x=-w_{xx} . 
\end{equation}
This yields Burgers' equation 
\begin{equation}\label{nullburgers}
v_t + v v_x +v_{xx} =0
\end{equation}
for $u=\int v\; dx$, 
or equivalently 
\begin{equation}\label{nullpotburgers}
w_t +\tfrac{1}{2}w_x{}^2 +w_{xx}=0
\end{equation}
in potential form, for $v=\int w\; dx$. 
We recall that, from \secref{2dimnullcase}, 
Burgers' equation has a gradient-energy structure, 
\begin{equation}
v_t=-D_x(\exp(w) D_x\delta \mk{H}/\delta v)
\end{equation}
and in potential form 
\begin{equation}
w_t=\exp(w) \delta \mk{H}/\delta w
\end{equation}
with $\mk{H}=\txtint H\;dx$ where 
\begin{equation}
H = \tfrac{1}{2}\exp(-w)v^2 
\end{equation}
is the energy density. 
Next, consider the flow generated by 
\begin{equation}\label{nullkdvflow}
h_+=-v_{xx}=-w_{xxx} . 
\end{equation}
This yields the KdV equation 
\begin{equation}\label{nullkdv}
v_t + v v_x +v_{xxx} =0
\end{equation}
for $u=\int v\; dx$, 
or equivalently in potential form, 
\begin{equation}\label{nullpotkdv}
w_t + \tfrac{1}{2}w_x{}^2 +w_{xxx} =0
\end{equation}
for $v=\int w\; dx$. 
The KdV equation has the Hamiltonian structure, 
\begin{equation}
v_t=\Hop(\mk{H}/\delta v)
\end{equation}
with $\mk{H}=\txtint H\;dx$ 
where 
\begin{equation}
H = \tfrac{1}{2}v_x{}^2 -\tfrac{1}{6}v^3
\end{equation}
is the Hamiltonian density,
and where
\begin{equation}\label{nullHop}
\Hop=D_x
\end{equation}
is a Hamiltonian operator. 
Equivalently, 
this structure has the potential form 
\begin{equation}
w_t=-\Dop(\delta \mk{H}/\delta w)
\end{equation}
where 
\begin{equation}
\Dop=D_x^{-1}
\end{equation}
is a Hamiltonian operator. 

Both Burgers' equation \eqref{nullburgers} and the KdV equation \eqref{nullkdv}
are integrable systems. 
However, their respective recursion operators do not appear 
in the underlying $x$-translation invariant system 
\eqref{nullvt}--\eqref{nullvpotential}, 
or \eqref{nullwt}--\eqref{nullwpotential}, 
describing the class of flows \eqref{nullut} with $a=0$ and $b\neq0$. 

Finally, we consider the class of flows with $a=b=c=d=0$, 
\begin{equation}\label{nulllinearwt}
w_t = h_+, 
\quad
w_x=v , 
\end{equation}
or equivalently 
\begin{equation}\label{nulllinearvt}
v_t = D_x h_+,
\quad
v_x=u . 
\end{equation}
These flows give rise to a hierarchy of linear flows 
generated by a recursion operator $\Rop=D_x$,
starting from the root flow $h_+=v_x=w_{xx}$. 
Hence we have the following result. 

\begin{thm}
\label{nulllinhierarchy}
There is a hierarchy of integrable linear flows on $v(t,x)$ given by 
\begin{align}
v_t = h_{+}^{(n)} & 
= \Rop^n(v_x), 
\quad
n=1,2,\ldots 
\label{nulllinflows}
\\& 
=\begin{cases} 
-\delta \mk{H}^{(l)}/\delta v, 
\quad
l=(n+1)/2, &
n=1,3,\ldots
\\ 
-D_x(\delta \mathfrak{H}^{(l)}/\delta v), 
\quad
l=n/2, & 
n=2,4,\ldots
\end{cases} 
\label{nulllinHamilflows}
\end{align}
in terms of energies/Hamiltonians $\mk{H}^{(l)}=\int H^{(l)} dx$, 
where
\begin{equation}
H^{(l)}=
\tfrac{1}{2} (-1)^{l-1} (D_x^{l} v)^2
\quad 
l=1,2,\ldots
\label{nulllinHam}
\end{equation}
are local densities,
and where
\begin{equation}
\Rop=D_x
\end{equation}
is a recursion operator. 
\end{thm}

This theorem is a direct counterpart of the hierarchy of 
linear flows \eqref{2dimhflows} shown in \thmref{2dimhuhierarchy}
for the case of null curves in 2 dimensions. 

The $+1$ flow in the hierarchy \eqref{nulllinflows} 
is given by the heat equation
\begin{equation}\label{nullheat}
v_t=v_{xx}
\end{equation}
which has the gradient structure
\begin{equation}
v_t=-\delta \mk{H}^{(1)}/\delta v
\end{equation}
in terms of the energy density
\begin{equation}\label{nullH1}
H^{(1)} = \tfrac{1}{2} v_x{}^2 . 
\end{equation}
All of the odd flows in this hierarchy have a similar structure. 

The $+2$ flow in the hierarchy \eqref{nulllinflows} 
is given by 
\begin{equation}\label{nullairy}
v_t=v_{xxx}
\end{equation}
which is the Airy equation. 
This equation has the Hamiltonian structure
\begin{equation}
v_t=-\Hop(\delta \mk{H}^{(1)}/\delta v)
\end{equation}
where the Hamiltonian $\mk{H}^{(1)}$ is the same expression 
as the energy integral appearing in the $+1$ flow,
and where $\Hop$ is the Hamiltonian operator \eqref{nullHop}. 
There is a similar structure for all of the even flows 
in the hierarchy. 

To conclude, we now work out the underlying equations of motion 
for the curve flows determined by the preceding flows. 
We begin by using the Frenet relations \eqref{nullframe}
to express the frame vectors \eqref{nulladaptedframe} 
in terms of the tangent vector $\pos_x$:
\begin{align}
& e_\parallel = \pos_x
\\
& \tilde{e}_+ = u^{-1} e_\parallel{}_x = u^{-1} \pos_{xx}
\\
& \tilde{e}_- = u \Nop(e_\parallel{}_x) = u \Nop(\pos_{xx})
\end{align}
where, recall, $\Nop$ is the linear map on null vectors
in the Minkowski plane orthogonal to $e_\parallel$ in $\Mink{2}$. 
In particular, 
for any null vector $v$ such that $\eta(e_\parallel,v)=0$, 
this map is uniquely determined by the properties 
\begin{equation}
\eta(\Nop(\vec v),e_\parallel) = 0,
\quad
\eta(\Nop(\vec v),\vec v) = -1,
\quad
\eta(\Nop(\vec v),\Nop(\vec v))=0 . 
\end{equation}
We next derive an expression for the curve invariant $u$ 
in terms of $x$-derivatives of these frame vectors. 
From the Frenet equations \eqref{nullEx} of the frame, 
we have 
$e_\parallel{}_{xx} = u_x \tilde{e}_+ = u_x u^{-1} e_\parallel{}_x$. 
Then, since $e_\parallel{}_x$ is a null vector orthogonal to $e_\parallel$, 
we obtain 
\begin{equation}
(\ln u)_x = -\eta(e_\parallel{}_{xx}, \Nop(e_\parallel{}_x))
= -\eta(\pos_{xxx}, \Nop(\pos_{xx}))
\end{equation}
and hence
\begin{equation}\label{nullu}
u = \txtint \exp(-\eta(\pos_{xxx}, \Nop(\pos_{xx}))) dx . 
\end{equation}
The flow vector \eqref{nullflow} now leads to the following results. 

\begin{prop}
The flows \eqref{nullburgersflow} and \eqref{nullkdvflow}
producing, respectively,   
Burgers' equation \eqref{nullburgers} and the KdV equation \eqref{nullkdv}
determine an inelastic spacelike curve flow with a null principal normal, 
where the components of the flow vector \eqref{nullflow} are given by 
\begin{gather}
h_\parallel = -v=-\txtint u\;dx,
\quad
h_{-} = 1
\\
h_{+} = -u
\quad\text{and}\quad
h_{+} = -u_x
\end{gather}
where $u$ is the curve covariant \eqref{nullu} and $v$ is its potential. 
\end{prop}

The resulting curve flows are explicitly given by 
\begin{equation}\label{nullburgerseom}
\pos_t=-v\pos_x + u\Nop(\pos_{xx}) -\pos_{xx},
\quad 
\eta(\pos_x,\pos_x)=1,
\quad
\eta(\pos_{xx},\pos_{xx})= 0
\end{equation}
which corresponds to Burgers' equation \eqref{nullburgers},
and 
\begin{equation}\label{nullkdveom}
\pos_t=-v\pos_x + u\Nop(\pos_{xx}) -\pos_{xxx},
\quad 
\eta(\pos_x,\pos_x)=1,
\quad
\eta(\pos_{xx},\pos_{xx})= 0
\end{equation}
which corresponds to the KdV equation \eqref{nullkdv}. 

\begin{prop}
In the hierarchy \eqref{nulllinflows}, 
each flow determines an inelastic spacelike curve flow with a null principal normal, 
where the components of the flow vector \eqref{nullflow} are given by 
\begin{equation}
h_\parallel = 0, 
\quad
h_{-} = 0,
\quad
h_{+}^{(n)} = \Rop^n(v_x) = D_x{}^n u,
\quad
n=1,2,\ldots
\end{equation}
in terms of the curve covariant \eqref{nullu}. 
\end{prop}

The curve flow $n=1$ 
corresponding to the heat equation \eqref{nullheat} is explicitly given by 
\begin{equation}\label{nullheateom}
\pos_{t}=\pos_{xx},
\quad 
\eta(\pos_x,\pos_x)=1,
\quad
\eta(\pos_{xx},\pos_{xx})= 0
\end{equation}
which is a non-stretching variant of the heat map equation, 
in Minkowski space $\Mink{2}$. 
Similarly, 
the curve flow $n=2$ corresponding to the Airy equation \eqref{nullairy} 
is given by 
\begin{equation}\label{nullairyeom}
\pos_{t} = \pos_{xxx}
\quad 
\eta(\pos_x,\pos_x)=1, 
\quad
\eta(\pos_{xx},\pos_{xx})\neq 0
\end{equation}
which is a non-stretching Airy map equation
in Minkowski space $\Mink{2}$. 

Note that all of these equations of motion 
\eqref{nullburgerseom}, \eqref{nullkdveom}, 
\eqref{nullheateom}, \eqref{nullairyeom} 
are invariant under the isometry group $ISO(2,1)$ of $\Mink{2}$ 
and thus describe geometric non-stretching motions of the curve.

\section{Conclusion}
\label{conclude}

There are several interesting directions in which the methods and results
in this paper can be extended. 

Firstly, 
one can consider inelastic curve flows in $n$-dimensional Minkowski space 
$\Mink{n}$ for $n\geq 4$. 
Inelastic flows of timelike curves will yield 
an integrable multi-component $SO(n-1)$-invariant vector version of 
the defocusing mKdV equation and its bi-Hamiltonian integrability structure,
while spacelike curves will similarly yield a 
$SO(n-2,1)$-invariant variant of that equation. 
Inelastic flows of null curves will yield multi-component Burgers' equations
and their symmetry-integrability structures. 

Secondly, 
one can consider inelastic curve flows in Lorentzian symmetry spaces 
$M=G/SO(n-1,1)$, with $G=SO(n,1)$ and $G=SU(n-1,1)$ for $n\geq 3$. 
These two spaces are curved generalizations of Minkowski space 
in which their frame bundles have the same $SO(n-1,1)$ gauge group 
of rotations and boosts as in the flat space $\Mink{n}$. 
The space $M=SO(n,1)/SO(n-1,1)$ will yield the same 
multi-component vector defocusing mKdV equations obtained from $\Mink{n}$,
whereas the space $M=SU(n-1,1)/SO(n-1,1)$ will yield 
different multi-component vector mKdV equations,
analogously to the geometric origin of the two known 
multi-component vector focusing mKdV equations from \cite{Anc06}
the Riemannian symmetric spaces 
$M=G/SO(n)$, with $G=SO(n+1)$ and $G=SU(n)$ for $n\geq 3$. 
For $n=3$, the Lorentzian symmetric space $M=SO(3,1)/SO(2,1)$ 
will yield the same two integrable defocusing NLS equations
as found in this paper,
while the space $M=SU(2,1)/SO(2,1)$ 
will yield two other integrable defocusing NLS equations 
(and their bi-Hamiltonian integrability structures). 

Lastly, 
one can also obtain integrable sinh-Gordon equations,
including (split) complex versions and multi-component vector versions, 
by considering inelastic curve flows related to non-stretching wave map 
equations in $M=SO(n,1)/SO(n-1,1)$ and $M=SU(n-1,1)/SO(n-1,1)$,
in analogy with the various versions of sine-Gordon equations 
obtained from \cite{Anc06,Anc08} the Riemannian symmetric spaces 
$M=SO(n+1)/SO(n)$ and $M=SU(n)/SO(n)$,
for $n\geq 3$.

\end{document}